\documentclass[11pt]{article}
\usepackage[english,activeacute]{babel}
\usepackage{natbib}
\usepackage{comment}
\usepackage{float}
\usepackage[hidelinks]{hyperref}
\usepackage{mathrsfs}
\usepackage{enumitem}
\usepackage[font={small,it}]{caption}
\usepackage{amsmath,amsfonts,amsthm,amssymb}
\usepackage{bm,rotating,multirow,dsfont,graphicx}
\usepackage[dvipsnames]{color}
\usepackage{url}
\usepackage{multicol}
\usepackage{multirow}
\usepackage[T1]{fontenc}
\usepackage{flafter}
\usepackage{appendix}
\usepackage{subfigure}
\usepackage{xcolor}
\usepackage{ragged2e}
\usepackage{listings}
\usepackage{authblk}

\makeatletter
\def\hlinewd#1{%
	\noalign{\ifnum0=`}\fi\hrule \@height #1 %
	\futurelet\reserved@a\@xhline}
\makeatother
\usepackage[all,cmtip]{xy}

\def\exp#1{{\rm exp}{#1}}
\def\frac#1#2{{{#1}\over{#2}}}

\def\@roman#1{\romannumeral #1}

\begin{document}

	\def\spacingset#1{\renewcommand{\baselinestretch}{#1}\small\normalsize}\spacingset{1}
	
	\title{Leadership and Engagement Dynamics in Legislative Twitter Networks: Statistical Analysis and Modeling}

	\author{
		Carolina Luque$^{1}$\footnote{Contact: cluque2.d@universidadean.edu.co; cmluquez@unal.edu.co} \\
		Juan Sosa$^2$\footnote{Contact: jcsosam@unal.edu.co }
	}
	
	\date{%
		$^{1}$Universidad Ean, Bogotá, Colombia.\\%
		$^{2}$Universidad Nacional de Colombia, Bogotá, Colombia\\%
	}
	
	\maketitle
	
	\begin{abstract}
		\noindent
		In this manuscript, we analyze the interaction network on Twitter among members of the 117th U.S. Congress to assess the visibility of political leaders and explore how systemic properties and node attributes influence the formation of legislative connections. 
		We employ descriptive social network statistical methods, the exponential random graph model (ERGM), and the stochastic block model (SBM) to evaluate the relative impact of network systemic properties, as well as institutional and personal traits, on the generation of online relationships among legislators.
		Our findings reveal that legislative networks on social media platforms like Twitter tend to reinforce the leadership of dominant political actors rather than diminishing their influence. 
		However, we identify that these leadership roles can manifest in various forms. Additionally, we highlight that online connections within legislative networks are influenced by both the systemic properties of the network and institutional characteristics.
	\end{abstract}

	\noindent
	{Keywords: Exponential random graph model; Legislative online networks; Social networks; Stochastic block model; Twitter.}

	\spacingset{1.1} 

	\section{Introduction}

	Twitter (currently $\mathbb{X}$) is the preferred political mobilization channel for political actors \citep{golbeck2010twitter, mankad2015analysis, himelboim2017classifying, van2020twitter, praet2021patterns} and one of the most relevant sources of information for investigating their corresponding ways of online interaction \citep{vergeer2015twitter, wojcik2019legislative}. 
	Previous studies highlight the use of Twitter data and the relevance of social network statistical theory to provide empirical evidence on political polarization \citep{del2018echo, praet2021patterns}, opinion leadership \citep{borge2017opinion}, political group structure \citep{cherepnalkoski2016retweet}, public engagement \citep{grant2010digital}, among others. 
	However, these studies have mainly focused on explaining how connections between political actors emerge from a partisan perspective, without thoroughly examining the importance of other institutional traits, personal attributes, and systematic network properties as determining factors in relationship-building and leadership positioning within the context of online legislative networks.

	Previous research indicates that for scholars in the social sciences, it is challenging to understand how social cohesion emerges and how leadership and hierarchies are visualized in the chambers of popular sovereignty regarding an online environment \citep[][]{grant2010digital, borge2017opinion, wojcik2019legislative}.
	In order to make a significant contribution in this direction, we propose analyzing the configuration of the interaction network on Twitter among the members of the 117th U.S. Congress to assess the visibility of political leaders and explore the influence of systemic properties and nodal attributes in the formation of legislative ties.
	This dataset provides a snapshot of contemporary online political participation and is suitable for studying the diffusion of information in social networks \citep[][]{fink2023congressional}.
	Consequently, we examine two hypotheses:
	\begin{quote}
		\textbf{Hypothesis 1:} The configuration of legislative networks on social media platforms like Twitter favors the participation of already visible political leaders rather than highlighting new ones.
	\end{quote}
	\begin{quote}
		\textbf{Hypothesis 2:} The ties in online legislative networks are more influenced by the systemic properties of the network than by the institutional or personal traits of its members.    
	\end{quote}

	We validate the first hypothesis by analyzing the network topology \citep[][]{kirkland2014measurement,li2021review}. 
	In other words, we calculated descriptive metrics (structural characteristics or systematic properties of the network) to characterize, from a relational perspective, those actors who play a prominent role in shaping the flow of information. 
	In this sense, we associate leadership with a prominent role in the dissemination of information to others, and we measure it using actor-specific social network metrics. 
	Unlike other authors who examine this hypothesis, we incorporate additional metrics beyond the degree to characterize individual centrality \citep[e.g.,][]{borge2017opinion}. Additionally, we identify roles within the network based on the calculated statistics.

	Regarding the second hypothesis, we implement both an exponential random graph model \citep[ERGM;][]{frank1986markov, wasserman1996logit, morris2008specification, lusher2013exponential} and a stochastic block model \citep[SBM;][]{abbe2018community,lee2019review}. 
	We do so to evaluate the relative influence of the systematic properties of the network (endogenous variables) as well as institutional and personal traits (exogenous variables) in generating online relationships between legislators. 
	Our modeling strategy also allows us to analyze which of these variables contribute to the formation of underlying communities within the network and to recognize the leadership positioning of legislative groups. 
	As institutional traits, we consider party affiliation and membership in the legislative chamber. 
	The personal attributes include race, ethnicity, religion, gender, age, length of service in Congress, and membership in the LGBTIQ+ community. 
	The latter, as far as we know, is an attribute that has not been widely studied as a homophily effect in online legislative networks. 
	Recent studies advocate for the inclusion of this trait in political science research \citep[e.g.,][]{ayoub2022not}.

	Exploring ways in which politicians connect with each other, considering institutional, personal, and structural factors, is valuable for enhancing our understanding of political participation phenomena outside the traditional legislative arena \citep[][]{cook2016american}. 
	Such exploration is crucial for uncovering informal patterns of influence, collaboration, and representation, particularly as digital platforms play a growing role in shaping political dynamics and public engagement. 
	Analyzing these broader connections can reveal insights into political alliances and diverse forms of representation that go beyond party lines and formal structures
	\citep{grant2010digital,cook2016american,borge2017opinion}.
	This is why we contribute to the analysis of political participation of parliamentary elites on social media platforms like Twitter by adding a perspective that covers relational factors beyond partisanship, an institutional trait that has predominated in previous research in the field \citep[][]{grant2010digital,cook2016american,borge2017opinion,cherepnalkoski2016retweet,del2018echo,praet2021patterns}.

	The document is structured as follows. 
	In Section \ref{sec:congress}, we briefly present the structure and general context of the 117th U.S. Congress. 
	In Section \ref{sec:methodology}, we describe the methodological aspects of the study. 
	In Section \ref{sec:results}, we present the research results. Specifically, we characterize the prominent actors, connectivity, and information flow within the system. Additionally, we examine the exogenous and endogenous variables that influence the network’s structure and the formation of online political communities. 
	Finally, in Section \ref{sec:discussion}, we discuss our main findings and propose future research directions.

	\section{117th U.S. Congress} \label{sec:congress}

	The 117th U.S. Congress serves from January 3, 2021, to January 3, 2023. 
	It is a bicameral body composed of the House of Representatives (lower chamber) and the Senate (upper chamber). 
	The lower chamber has 435 members, with seats allocated among the 50 states based on population, according to congressional reapportionment \citep[][]{rossiter2018congressional}. 
	The majority of seats in this chamber are held by the Democrat Party, led by \textit{Nancy Pelosi} as Speaker, who is responsible for presiding over sessions and setting the legislative agenda.
	The majority leader
	of the Democrats is \textit{Steny Hoyer}, and the whip
	is \textit{James Clyburn}.
	The Republican minority leader is \textit{Kevin McCarthy}, and the minority whip is \textit{Steve Scalise}.

	The Senate is composed of 100 legislators, with two senators representing each of the 50 states, regardless of population. Initially, the upper chamber is dominated by Republicans, but the senate control shifts after President \textit{Joe Biden} takes office. 
	His Vice President, \textit{Kamala Harris}, serving as President of the Senate, holds the power to cast tie-breaking votes, giving the Democrats the majority in cases of parity. 
	\textit{Chuck Schumer} acts as the Democrat majority leader in this chamber, while \textit{Mitch McConnell} leads the Republican minority.
	The party whips in the Senate are \textit{Dick Durbin} for the Democrats and \textit{John Thune} for the Republicans. This Congress also includes 6 non-voting delegates \citep[for more details, see][]{mamet2021representation}.

	The 117th U.S. Congress is characterized by intense political activity and polarization \citep[][]{binder2022,bond2024contemporary}. 
	Among its most significant legislative events is the impeachment trial of former Republican President Donald Trump for inciting insurrection following the attack on the Capitol on January 6, 2021. 
	Although Trump was acquitted, this event is considered one of the most important in recent U.S. political history \citep[][]{pearson2022legacies}. 
	Additionally, this Congress managed significant legislative responses to the COVID-19 pandemic, including economic stimulus packages for those affected \citep[][]{clemens2021politics,bivens2023federal}. It also passed a historic infrastructure law for the modernization and economic growth of the country \citep[][]{zhang2022review}, and promoted the implementation of climate policies \citep[][]{lawson2021clean}.

	\section{Methodology} \label{sec:methodology}

	In this section, we address all the essential factors required to conduct a thorough analysis, including data collection and analysis methods. 
	We also discuss the rationale behind our approach, ensuring that the analysis is aligned with our research objectives.

	\subsection{Data collection and network structure}

	We used interaction data from members of the 117th U.S. Congress on the social media platform Twitter, covering the period from February 9, 2022, to June 9, 2022 \citep{fink2023congressional}. 
	These data are available on Zenodo\footnote{\scriptsize \url{https://doi.org/10.5281/zenodo.8253486}, \url{https://zenodo.org/record/8253486}} and SNAP\footnote{\scriptsize \url{https://snap.stanford.edu/data/congress-twitter.html}} (Stanford Network Analysis Project). Additionally, we included node-level variables\footnote{\scriptsize Sources consulted: \url{https://www.senate.gov/senators/index.htm}, \url{https://history.house.gov/Congressional-Overview/Profiles/117th/}, and \url{https://en.wikipedia.org/wiki/117th_United_States_Congress}} such as party affiliation, race, ethnicity, religion, gender, legislative chamber, LGBTIQ+ community membership, age, and years of service.
	Table \ref{tab:descrip} presents the percentage distribution of the qualitative variables. 
	We have legislators representing all 50 states, with ages ranging from 29 to 91 years, an average age of 62.33, and a standard deviation of 11.82. The years of service among members of Congress range from 2 to 48 years, with an average of 10.90 years and a standard deviation of 9.51.

	\begin{table}[!htb]
		\scriptsize
		\centering
		\setlength{\tabcolsep}{2pt} 
		\begin{tabular}{llc}
			\hline
			\multicolumn{1}{c}{\textbf{Variable}} &                                                                                                                         & \textbf{Proportion} \\ \hline
			\textit{Party}                      & \textit{Democrat}                                                                                                      & 0.522               \\
			& \textit{Republican}                                                                                                    & 0.470               \\
			& \textit{Independent}                                                                                                  & 0.008               \\ \hline
			\textit{Chamber}                       & \textit{House of Representatives (lower chamber)}                                                                                & 0.806               \\
			\textit{}                             & \textit{Senate (upper chamber)}                                                                                                  & 0.194               \\ \hline
			\textit{Race}                         & \textit{White}                                                                                                         & 0.750               \\
			& \textit{Black}                                                                                                          & 0.126               \\
			& \textit{Other}                                                                             & 0.084               \\
			& \textit{Asiatic}                                                                                                       & 0.032               \\
			& \textit{Native American}                                                                                               & 0.008               \\ \hline
			\textit{Ethnicity}                & \textit{No Hispanic}                                                                                                     & 0.920               \\
			& \textit{Hispanic}                                                                                                        & 0.080               \\ \hline
			\textit{Religion}                     & \textit{\begin{tabular}[c]{@{}l@{}}Christian\\ (Protestant, Catholic, Roman Catholic, Evangelical, Mormon, etc.)\end{tabular}}  & 0.861               \\
			& \textit{\begin{tabular}[c]{@{}l@{}}Other\\ (Buddhist, Hindu, Muslim, Jewish, Humanist, Unaffiliated, etc.)\end{tabular}} & 0.139               \\ \hline
			\textit{Sex}                         & \textit{Male}                                                                                                      & 0.701               \\
			\textit{}                             & \textit{Female}                                                                                                       & 0.299               \\ \hline
			\textit{LGBTIQ+}            & \textit{No}                                                                                                             & 0.981               \\
			& \textit{Yes}                                                                                                             & 0.019               \\ \hline
		\end{tabular}
		\caption{Percentage distribution of qualitative node-level variables.}
		\label{tab:descrip}
	\end{table}

	The choice of this dataset is motivated by the dominant position of the United States in international politics and political science research \citep{praet2021patterns}. Studies on the U.S. Congress serve as an inspiration for future research on deliberative bodies with limited empirical evidence \citep[see][]{gamm2002legislatures,luque2022operationalizing}.

	In this way, we define a directed and weighted network \(G = (V, E)\), where each node \(i \in V\) represents a political actor in the network and \(e = \{i, j\} \in E\), with \(i, j \in V\), represents an edge (connection) between pairs of nodes. 
	Specifically, \(i\) and \(j\) in \(V\) are connected on Twitter if one of them retweeted, quoted, replied to, or mentioned the other.
	Furthermore, \(\{i, j\} \neq \{j, i\}\) for all \(i, j \in V\). 
	Thus, we have a network of 475 nodes and 13,289 edges.

	Although the 117th U.S. Congress comprises 535 members and 6 delegates, we only include those who have an active Twitter account and issued at least 100 tweets at the time of data collection \citep[for more details, see][]{fink2023congressional}. 
	The weights \(w_{i, j}\) of the network are empirical probabilities of influence between pairs of Congress members \citep[see][]{fink2023congressional}. 
	The weight \(w_{i, j}\) is the ratio of the total interactions from \(i\) to \(j\) to the total number of tweets issued by Congress member \(i\) during the observation period. 
	The weights \(w_{i, j}\) range from approximately 0.0005 to 0.1306. 
	The exclusion of Congress members who issued fewer than 100 tweets implies that \(0 < w_{i} = \sum_{j: j \in e = \{i, j\}} w_{i, j} < 1\).

	\subsection{Methods}

	\subsubsection{Network topology}

	In the same spirit as \cite{ringe2016pinpointing}, we implement centrality measures to characterize the actors in the network and identify those who play a predominant role in the configuration of information within the system. 
	In this regard, we evaluate metrics such as \textit{degree (in/out)}, \textit{strength (out)}, \textit{closeness centrality}, \textit{betweenness centrality}, and \textit{eigen centrality} \citep[for more details, see ][]{ward2011network, kolaczyk2014statistical}. 
	Additionally, we implement the HITS algorithm \citep[Hyperlink-Induced Topic Search;][]{kleinberg1999authoritative}, available in the \texttt{igraph} package in \texttt{R} \citep{csardi2013package}, to recognize highly referenced (\textit{authorities}) and connected or sociable (\textit{hubs}) actors within the network.

	Furthermore, we analyzed the network's connectivity employing several metrics such as \textit{density}, \textit{transitivity}, \textit{reciprocity}, \textit{cliques}\footnote{\scriptsize Clans or subsets of nodes where every pair of vertices is connected in some direction \citep{kolaczyk2014statistical}.}, \textit{local} and \textit{global clustering coefficient} \citep[for more details, see ][]{rakhmawati2020social, li2021review}. 
	Also, we examine the flow of information in the network by means of the \textit{assortativity} \citep[selective mixing or homophily;][]{newman2002assortative, noldus2015assortativity} to determine the tendency of actors to connect with others who are similar or dissimilar regarding the node and structural characteristics under consideration \citep{schwarzenbach2024extremists}. 
	An assortativity value close to $-1$ indicates that nodes tend to connect with others that differ in a specific attribute, while a value close to $1$ suggests that nodes are more likely to connect with others sharing the same characteristic. A value near $0$ indicates little to no correlation between connections and node attributes.

	\subsubsection{Exponential Random Graph Model}

	We implement an ERGM \citep[][]{frank1986markov, wasserman1996logit, morris2008specification, lusher2013exponential} to examine the effect of exogenous and endogenous variables on the formation of links in the 117th U.S. Congress network. 
	This approach allows us to model network-generating processes, both individual and structural, which are common in political science \citep{cranmer2011inferential}.

	The probabilistic representation of the ERGM is given by $p(\mathbf{y}\mid\boldsymbol{\theta}) = \tfrac{1}{\kappa}\,\exp{\left\{ \boldsymbol{\theta}^{\textsf{T}}\boldsymbol{\mathsf{g}}(\mathbf{y}) \right\}}$, where $\mathbf{y}=[y_{i,j}]$ is a realization of the random adjacency matrix $\mathbf{Y}=[Y_{i,j}]$ associated with the 117th U.S. Congress network, $\boldsymbol{\theta} = [\theta_1,\ldots,\theta_K]^{\textsf{T}}$ is a $K$-dimensional vector of unknown model parameters, $\kappa\equiv\kappa(\boldsymbol{\theta})$ is the normalization constant, and $\boldsymbol{\mathsf{g}}(\mathbf{y}) = [\mathsf{g}_1(\mathbf{y}),\ldots,\mathsf{g}_K(\mathbf{y})]^{\textsf{T}}$ is a $K$-dimensional vector composed of network attributes (endogenous variables) or node attributes $\mathbf{x}$ (exogenous variables).

	Attributes are incorporated into the ERGM formulation as $\boldsymbol{\mathsf{g}}(\mathbf{y},\mathbf{x}) = \sum_{i<j} y_{i,j} h(\mathbf{x}_i, \mathbf{x}_j)$, where $\mathbf{x}_i$ represents the attribute vector of vertex $i$, and $h(\mathbf{x}_i, \mathbf{x}j)$ is a symmetric function that enables the model to include both main effects and second-order effects (homophily) related to observed characteristics. 
	Main effects are assessed using quantitative attributes, with $h(x_i, x_j) = x_i + x_j$, while homophily effects are evaluated through indicator variables in qualitative attributes, with $h(x_i, x_j) = I{{x_i = x_j}}$, or through differences in quantitative attributes, with $h(x_i, x_j) = |x_i - x_j|$. Thus, the model coefficients $\boldsymbol{\theta}$ capture the magnitude and direction of the effects of $\boldsymbol{\mathsf{g}}(\mathbf{y})$ on the probability of observing the network.

	The probability of observing the network $p(\mathbf{y}\mid\boldsymbol{\theta})$, can be re-expressed in logit scale in terms of the conditional probabilities of observing an edge between two actors while keeping the rest of the network fixed. 
	That is, $\text{logit},\textsf{Pr}(y_{i,j}=1\mid\mathbf{y}_{-(i,j)}) = \boldsymbol{\theta}^{\textsf{T}}\boldsymbol{\delta}_{i,j}(\mathbf{y})$, where $\mathbf{y}_{-(i,j)}$ corresponds to $\mathbf{y}$ except for the observation $y_{i,j}$, and $\boldsymbol{\delta}_{i,j}(\mathbf{y})$ is the change statistic that corresponds to the difference between the value of $\boldsymbol{\mathsf{g}}(\mathbf{y})$ when $y_{i,j}=1$ and $y_{i,j}=0$, while keeping the other values of $\mathbf{y}$ constant. This reparameterization is useful for interpreting the coefficients as the contribution of covariates to the probability (in logit scale) of observing a particular edge, conditioned on all other dyads remaining the same \citep{cranmer2011inferential}.

	We fit the model using the simulated maximum likelihood estimation (MLE) algorithm via Markov Chain Monte Carlo \citep[MCMC;][]{geyer1992constrained, hunter2006inference}. This estimation approach is available in the \texttt{ergm} package in \texttt{R} \citep{hunter2008ergm}. Additionally, we used the Akaike Information Criterion (AIC) and Bayesian Information Criterion (BIC) for model comparison as well as model selection \citep{chakrabarti2011aic}. Lower AIC and BIC values indicate a better balance between goodness of fit and model complexity.

	\subsubsection{Stochastic Block Model}

	We implement an SBM \citep[][]{abbe2018community, lee2019review} to capture underlying groupings in the network when the researcher has no prior knowledge about the number of classes (communities) or the class to which the vertices belong. In this sense, we assume that node $i\in V$ in the graph $G=(V,E)$ belongs to a single class within a partition $\mathcal{P} = {C_1,\ldots,C_Q}$ of $V$ with $Q$ communities.

	The formulation of the SBM is given by $p(\mathbf{y}\mid\boldsymbol{\theta}) = \tfrac{1}{\kappa}\,\exp{\left\{ \sum_{q,r}\theta_{q,r}\,L_{q,r}(\mathbf{y}) \right\}}$, where $\mathbf{y}=[y_{i,j}]$ is a realization of the random adjacency matrix $\mathbf{Y}=[Y_{i,j}]$ associated with the 117th U.S. Congress network, $L_{q,r}(\mathbf{y})$ is the number of edges connecting pairs of vertices from classes $q$ and $r$, $\pi_{q,r}$ is the interaction probability of a vertex from class $q$ with a vertex from class $r$, i.e., $\pi_{q,r}=\textsf{Pr}(y_{i,j}=1\mid i\in C_q, j\in C_r)$, $\boldsymbol{\theta}=(\theta_{1,1},\ldots,\theta_{Q,Q})$ is the vector of model parameters, and $\kappa\equiv\kappa(\boldsymbol{\theta})$ is the normalization constant.

	The membership of each vertex $i$ to a given class is determined independently according to a common probability distribution over the set $\{1,\ldots,Q\}$. 
	The latent random variable $z_{i,q} = 1$ if vertex $i$ belongs to community $q$ and $0$ otherwise. 
	Thus, $ \boldsymbol{z}_i\mid \boldsymbol{\alpha}\stackrel{\text{iid}}{\sim} \textsf{Categorical}(\boldsymbol{\alpha})$, which is equivalent to $\textsf{Pr}(z_{i,q} = 1\mid\alpha_q) = \alpha_q$, where $\boldsymbol{z}_i=(z_{i,1},\ldots,z_{i,Q})$, and $\boldsymbol{\alpha}=(\alpha_1,\ldots,\alpha_Q)$ with $\sum_{q=1}^Q \alpha_q = 1$. 
	Therefore, given the values of $\boldsymbol{z}_1,\ldots,\boldsymbol{z}_n$, dyads can be modeled as conditionally independent with a Bernoulli distribution as $y_{i,j}\mid\mathbf{\Pi},\boldsymbol{z}_i,\boldsymbol{z}_j \stackrel{\text{ind}}{\sim} \textsf{Bernoulli}\left( \pi_{\xi_i,\xi_j} \right)$, where $\mathbf{\Pi}=[\pi_{q,r}]$ is a $Q\times Q$ matrix containing the interaction probabilities, and $\xi_i=\xi(\boldsymbol{z}_i)$ denotes the position $q$ in $\boldsymbol{z}_i$ such that $z_{i,q} = 1$ (i.e., $\xi_i=q$ means that vertex $i$ belongs to community $q$).

	We fit the model using the variational expectation-maximization (EM) algorithm, available in the \texttt{blockmodels} package in \texttt{R} \citep{leger2016blockmodels, leger2022}. Additionally, we use the Integrated Classification Likelihood \citep[ICL;][]{biernacki1998assessing, lomet2012model} as a criterion for selecting the optimal number of communities. 
	Furthermore, we compare the resulting community assignments with the natural partitions derived from node variables. 
	To do this, we use clustering evaluation criteria such as the rand index \citep[rand; ][]{rand1971objective}, adjusted rand index \citep[adjusted.rand; ][]{hubert1985comparing}, and the normalized mutual information measure \citep[mni; ][]{danon2005comparing}.

	\section{Results} \label{sec:results}

	Here, we analyze the overall structure of the 117th Congress of the U. S. We identify that the corresponding network is a giant component, where each member is connected to at least one other individual within the legislative body in some direction. Additionally, we observe that the network lacks articulation points (i.e., critical actors whose absence would disrupt its connectivity), making it highly interconnected. However, its connectivity is not robust, as it is not possible to reach any political actor from any other individual through a directed walk.

	\subsection{Characterization of Network Actors} \label{sec:actores}

	We evaluate the in-degree (incoming connections) and out-degree (outgoing connections) of the network nodes, as well as the out-strength. 
	We do not consider the in-strength because influence probabilities (network weights) are calculated based on the emission rather than reception of links \citep[see][]{fink2023congressional}. 
	Both the degree and strength exhibit a similar distribution with positive skewness (as expected), which is typical of scale-free networks \citep[][]{bollobas2001degree}. The in-degree of actors in the network ranges from $0$ to $127$, while the out-degree ranges from $1$ to $210$. On average, each individual interacts with $28$ legislative members, with standard deviations of $21.993$ and $18.353$ for in-degree and out-degree, respectively. Additionally, approximately $48\%$ of the network actors have a degree (in/out) higher than the average.

	In Table \ref{tab:grado_fuerza}, we present the actors with the highest scores in these metrics. 
	We observe that the Republican minority leader \textit{Kevin McCarthy} and the Democrat majority leader \textit{Steny Hoyer} in the House of Representatives are endpoints for a considerable number of Congress members. 
	The number of different individuals who retweet, cite, reply to, or mention these representatives exceeds 100.
	Furthermore, we identify actors who do not play a role as information receivers in the system (i.e., individuals with an in-degree of zero). 
	For example, \textit{Chuck Grassley}, the Senate President pro tempore emeritus
	the Republican Representatives \textit{Lance Gooden} and \textit{Ann Wagner} from Texas and Missouri, respectively; and Democrat Representatives \textit{Josh Harder}, \textit{Tom O'Halleran}, \textit{Thomas Suozzi}, and \textit{Mark Warner} from California, Arizona, New York, and Virginia, respectively.

	\begin{table}[!htb]
		\scriptsize
		\centering
		\setlength{\tabcolsep}{2pt} 
		\begin{tabular}{clllcc}
			\cline{2-6}
			& \multicolumn{1}{c}{\textbf{Member}} & \multicolumn{1}{c}{\textbf{State}} & \multicolumn{1}{c}{\textbf{Party}} & \textbf{Chamber} & \textbf{Metric value} \\ \hline \\
			\multirow{5}{*}{\textbf{\begin{sideways} \begin{tabular}[c]{@{}l@{}} In- \\ degree  \end{tabular} \end{sideways} }} 
			& \textbf{Kevin McCarthy}              & California                          & Republican                          & Lower            & 127                    \\
			& Scott Franklin                       & Florida                             & Republican                          & Lower            & 121                    \\
			& Jeff Duncan                          & Carolina del sur                    & Republican                          & Lower            & 120                    \\
			& Donald Beyer                            & Virginia                            & Democrat                            & Lower            & 109                    \\
			& \textbf{Steny Hoyer}                 & Maryland                            & Democrat                            & Lower            & 108                    \\ \\
			\multirow{5}{*}{\textbf{\begin{sideways} \begin{tabular}[c]{@{}l@{}} Out- \\ degree  \end{tabular} \end{sideways} }}
			& \textbf{Nancy Pelosi}                & California                          & Democrat                            & Lower            & 210                    \\
			& \textbf{Kevin McCarthy}              & California                          & Republican                          & Lower            & 157                    \\
			& Bobby Rush                           & Illinois                            & Democrat                            & Lower            & 111                    \\
			& \textbf{Chuck Schumer}               & Nueva York                          & Democrat                            & Upper            & 97                     \\
			& \textbf{Steve Scalise}               & Luisiana                            & Republican                          & Lower            & 89                     \\ \\
			\multirow{5}{*}{\textbf{\begin{sideways} \begin{tabular}[c]{@{}l@{}} Out- \\ strength  \end{tabular} \end{sideways} }}        
			& \textbf{Nancy Pelosi}                & California                          & Democrat                            & Lower            & 0.944                  \\
			& \textbf{Steve Scalise}               & Luisiana                            & Republican                          & Lower            & 0.935                  \\
			& Bobby Rush                           & Illinois                            & Democrat                            & Lower            & 0.869                  \\
			& \textbf{Kevin McCarthy}              & California                          & Republican                          & Lower            & 0.827                  \\
			& Joe Wilson                           & Carolina del Sur                    & Republican                          & Lower            & 0.695                  \\ \\ \hline 
		\end{tabular}
		\caption{Top 5 congressional members by degree and strength. Members in bold have a specific role in the 117th Congress of the United States (see Section \ref{sec:congress}).} 
		\label{tab:grado_fuerza}
	\end{table}

	On the other hand, based on the out-degree, we identify popular information sources. 
	These are actors who establish connections with more than 18\% of the individuals in the system (see Table \ref{tab:grado_fuerza}). Among the prominent information sources are the Speaker of the House, \textit{Nancy Pelosi}, Republican leader \textit{Kevin McCarthy}, Republican whip in the House \textit{Steve Scalise}, and Democrat Majority Leader in the Senate \textit{Chuck Schumer}.
	Additionally, we identify actors who do not play a significant role as information sources in the system, as they connect with less than 1\% of the members. For example, \textit{Gregorio Kilili Camacho}, delegate from the Northern Mariana Islands in the House, Republican Representatives \textit{Claudia Tenney} and \textit{Kay Granger} from New York and Texas, respectively, and Democrat Representatives \textit{Ann Kirkpatrick} and \textit{Kurt Schrader} from Arizona and Oregon, respectively.

	Similarly, we examined the out-strength to identify the most influential actors in the system. This metric ranges from 0.002 to 0.944, with an average of 0.163 and a standard deviation of 0.126. Approximately 38.11\% of individuals have an out-strength greater than the average.
	Among the most influential actors are the Speaker of the House, \textit{Nancy Pelosi}, Republican whip \textit{Steve Scalise}, and Republican leader \textit{Kevin McCarthy}, each with an information transmission probability to system agents exceeding 80\% (see Table \ref{tab:grado_fuerza}).
	The least influential actors have an information transmission probability below 2.1\%. These members include \textit{Claudia Tenney}, and Democrats \textit{Gregorio Kilili Camacho}, \textit{Ann Kirkpatrick}, \textit{Lloyd Doggett}, and \textit{Dwight Evans}. The latter two are Representatives in the House from Texas and Pennsylvania, respectively.

\begin{figure}[!htb]
		\scriptsize
		\begin{center}
			\subfigure[In-degree.]{
				\centering
				\includegraphics[scale=0.35]{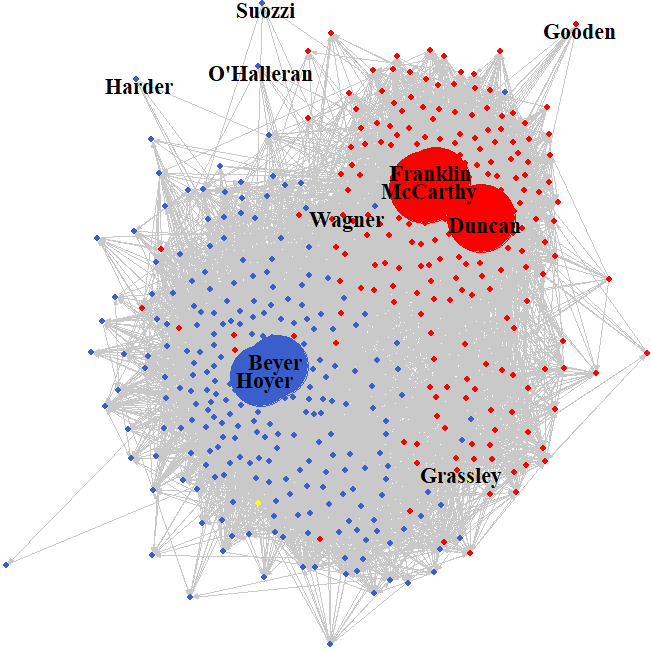}
				\label{fig:g_in}}
			\subfigure[Out-degree.]{
				\centering
				\includegraphics[scale=0.35]{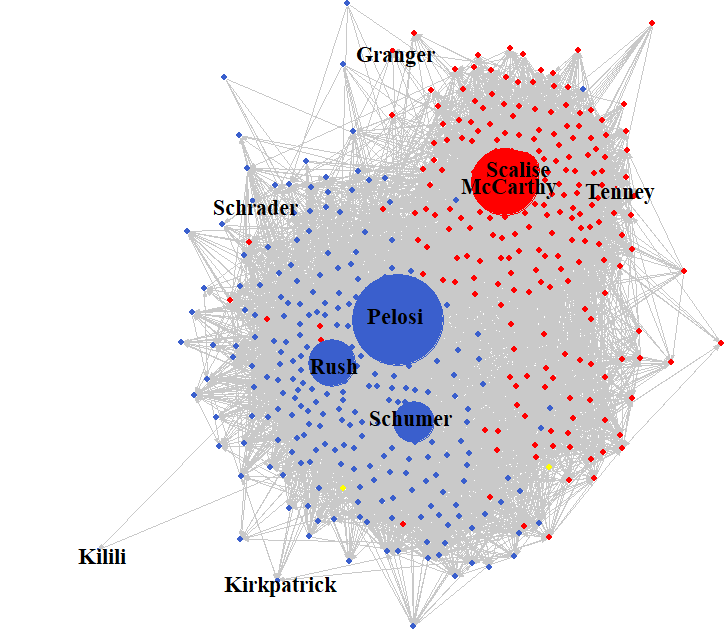}
				\label{fig:g_out}}     
			\subfigure[Out-strength.]{
				\centering
				\includegraphics[scale=0.35]{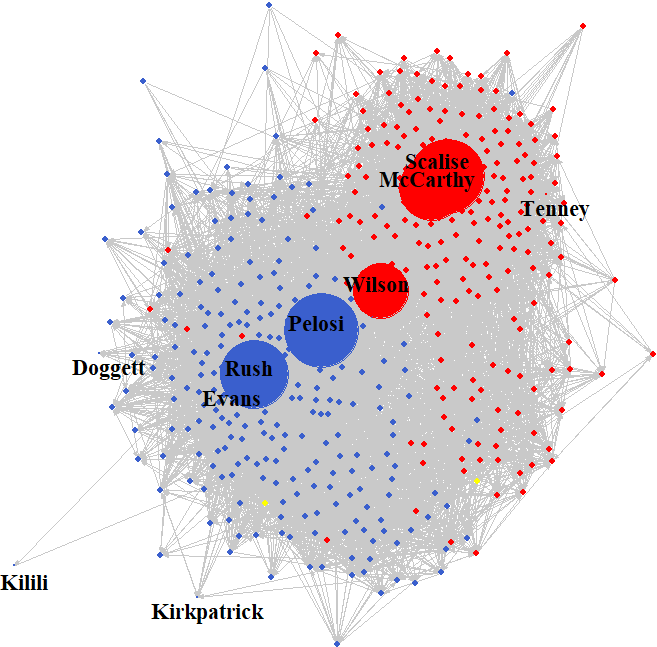}
				\label{fig:s_out}}
			\caption{117th Congress Twitter network of the U. S. In each graph, the actors with the highest and lowest scores in-degree and -strength metrics are highlighted. The size of the prominent nodes is proportional to the value of the metric, while the other nodes have a fixed size. Nodes representing members of the Republican Party are colored red, Democrat Party members are colored blue, and independents are colored yellow.}
			\label{fig:grado_fuerza}
		\end{center}    
	\end{figure}

	In Figure \ref{fig:grado_fuerza}, we observe that individuals who do not play a dominant role in the degree and strength metrics are located on the periphery of the system. In contrast, those who excel in these metrics are situated at the center. Specifically, these prominent actors are positioned where the majority of party members are concentrated.

\begin{figure}[!htb]
		\centering
		\scriptsize
		\begin{center}
			\subfigure[Closeness centrality.]{
				\centering
				\includegraphics[scale=0.35]{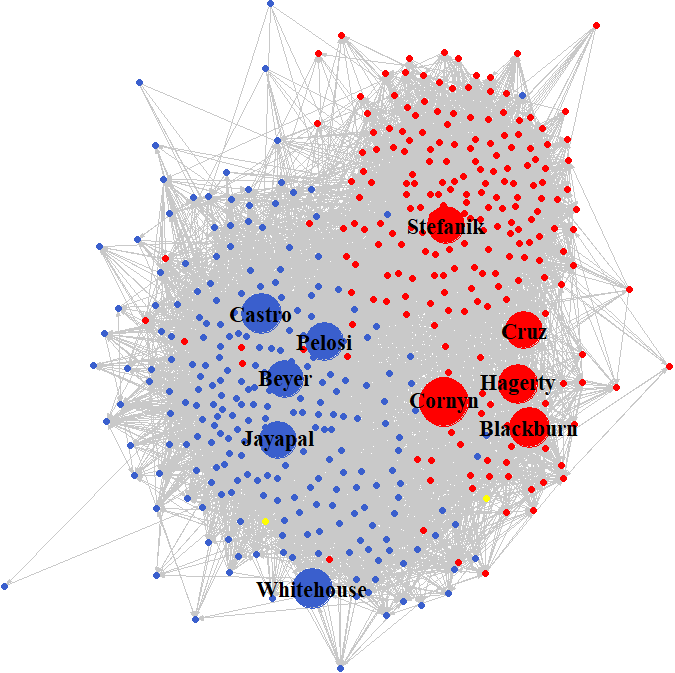}
				\label{fig:cc}}
			\subfigure[Betweenness centrality.]{
				\centering
				\includegraphics[scale=0.35]{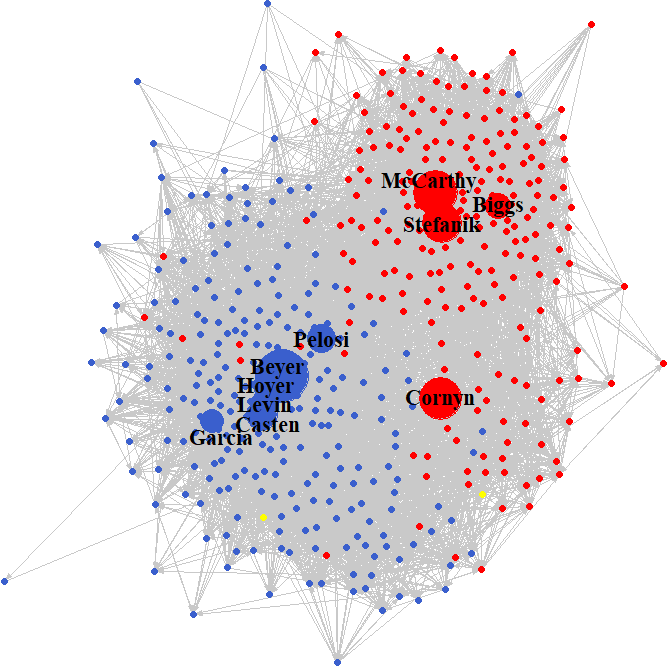}
				\label{fig:bc}}   
			
			\subfigure[Eigen centrality.]{
				\centering
				\includegraphics[scale=0.35]{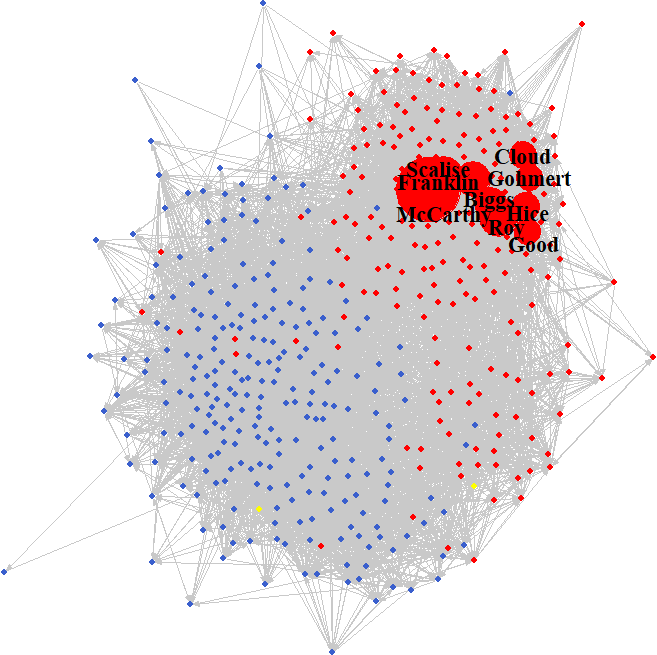}
				\label{fig:ec}}
			
			\subfigure[Authorities.]{
				\centering
				\includegraphics[scale=0.35]{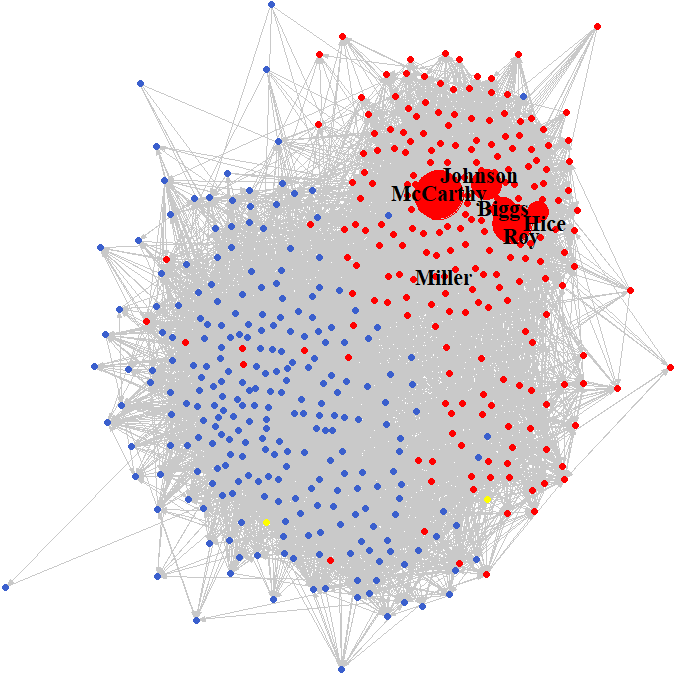}
				\label{fig:au}}
			\subfigure[Hubs.]{
				\centering
				\includegraphics[scale=0.35]{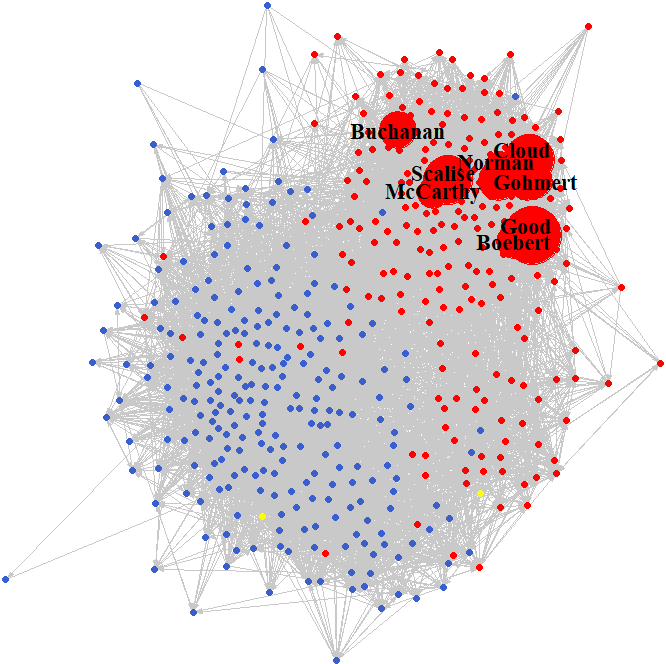}
				\label{fig:hub}}
			\caption{117th Congress Twitter network of the U. S. In each graph, the actors with the highest scores in centrality measures are highlighted. The size of these nodes is proportional to their metric values, while the other nodes remain fixed in size. In graphs (c)–(e), only some actors are visualized due to label overlap. The red color corresponds to members of the Republican Party, blue to the Democrat Party, and yellow to independents.} 
			\label{fig:centros}
		\end{center}
	\end{figure}

	Closeness centrality among network actors reveals that the inverse of the average geodesic distance between pairs of individuals ranges from 0.162 to 1.000. 
	The most prominent political figures based on their proximity to others in the system are \textit{John Cornyn} (1.000), \textit{Marsha Blackburn} (0.831), \textit{Sheldon Whitehouse} (0.828), \textit{Joaquín Castro} (0.819), \textit{Bill Hagerty} (0.796), \textit{Nancy Pelosi} (0.776), \textit{Ted Cruz} (0.770), \textit{Elise Stefanik} (0.764), \textit{Pramila Jayapal} (0.762), and \textit{Don Beyer} (0.759).
	In Figure \ref{fig:cc}, we highlight these legislators and demonstrate that they belong to the two major parties. Additionally, the positioning of members with high information diffusion capacity across the system shows a similar pattern for both political groups. 
	We note that actors with high closeness centrality who belong to the same chamber are closer to each other in the network. 
	For instance, Republican senators are more closely connected to each other and are further from \textit{Elise Stefanik}, a fellow Republican in the House of Representatives. Similarly, among Democrats, there is greater proximity among House members and a greater distance from \textit{Sheldon Whitehouse}, who is member of the Senate.

	The betweenness centrality of the nodes ranges from 0.000 to 0.085. This metric allows us to identify the key bridges in the network, which include \textit{Don Beyer} (0.085), \textit{Kevin McCarthy} (0.071), \textit{John Cornyn} (0.068), \textit{Elise Stefanik} (0.062), \textit{Sean Casten} (0.050), \textit{Nancy Pelosi} (0.045), \textit{Andy Levin} (0.044), \textit{Steny Hoyer} (0.041), \textit{Andy Biggs} (0.040), and \textit{Jesús García} (0.038). Among these, only \textit{Cornyn} is a Senator. 
	The rest are members of the House of Representatives.
	In Figure \ref{fig:bc}, we observe that both Democrats and Republicans have significant bridge actors who facilitate communication and information transfer within the system.

	On the other hand, node eigen centrality ranges from 0.001 to 1.000 and highlights influential centers, particularly among Republican House representatives: \textit{Kevin McCarthy} (1.000), \textit{Steve Scalise} (0.633), \textit{Chip Roy} (0.609), \textit{Mike Johnson} (0.562), \textit{Scott Franklin} (0.538), \textit{Andy Biggs} (0.505), \textit{Jody Hice} (0.454), \textit{Michael Cloud} (0.426), \textit{Louis Gohmert} (0.420), and \textit{Bob Good} (0.409). 
	In Figure \ref{fig:ec}, we locate some of these actors within the network. 
	Notably, none of the top 10 legislators with the highest eigen centrality are Senators or Democrats.

	Finally, we identify legislators who act as authorities and hubs in the 117th Congress Twitter network of the U.S. 
	Both metrics range from 0.000 to 1.000. The highly referenced legislators (authorities) include \textit{Kevin McCarthy} (1.000), \textit{Chip Roy} (0.825), \textit{Mike Johnson} (0.627), \textit{Andy Biggs} (0.503), \textit{Jody Hice} (0.447), \textit{Scott Franklin} (0.426), \textit{Byron Donalds} (0.409), \textit{Thomas Massie} (0.343), and \textit{Tom Emmer} (0.340). 
	On the other hand, the highly connected actors (hubs) who facilitate information flow in the system are \textit{Bob Good} (1.000), \textit{Michael Cloud} (0.859), \textit{Steve Scalise} (0.842), \textit{James Comer} (0.699), \textit{Vern Buchanan} (0.638), \textit{Louis Gohmert} (0.611), \textit{Lauren Boebert} (0.592), \textit{Tom Tiffany} (0.568), \textit{Ralph Norman} (0.555), and \textit{Kevin McCarthy} (0.546). 
	Our findings reveal that individuals with roles as authorities or hubs in the system are primarily from the Republican-controlled House of Representatives (see Figures \ref{fig:au} and \ref{fig:hub}).

	\subsection{Connectivity and Information Flow}

	Of all the possible connections in the system, we observe 5.9\% within the overall network structure. When broken down by party and chamber, we notice a higher density within specific groups. For the Republican Party, this metric rises to 10.5\%, while for the Democrat Party it decreases to 9.3\%. In the Senate, the density reaches 18.7\%, whereas in the House of Representatives it is 6.9\%.
	This suggests that information flow is more robust within natural groupings compared to the flow across the entire structure. However, in any scenario, the density remains below 20\%, indicating that, overall, information spreads slowly through the system due to limited communication pathways between political actors \citep[][]{friedkin1981development,li2021review}.

\begin{figure}[!htb]
		\centering
		\tiny
		\begin{center}
			\subfigure[Subgraph 1.]{
				\centering
				\includegraphics[scale=0.315]{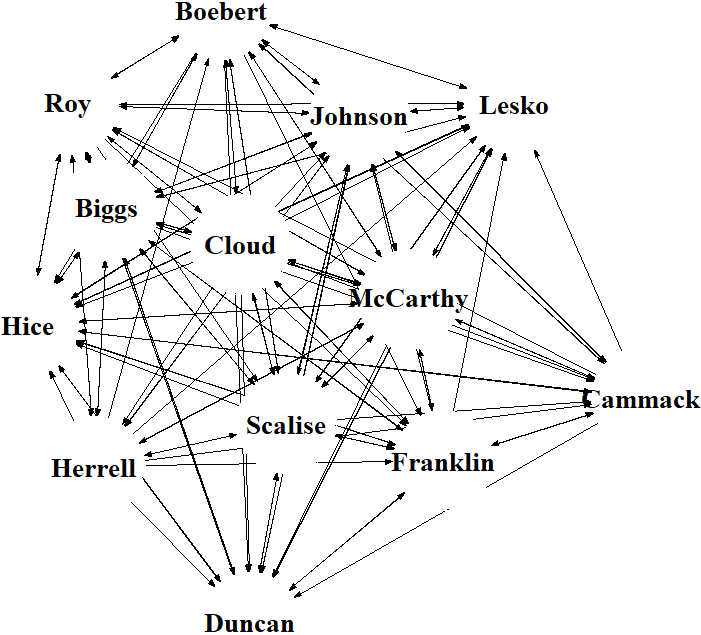}
				\label{fig:sub1}}
			\subfigure[Subgraph 2.]{
				\centering
				\includegraphics[scale=0.315]{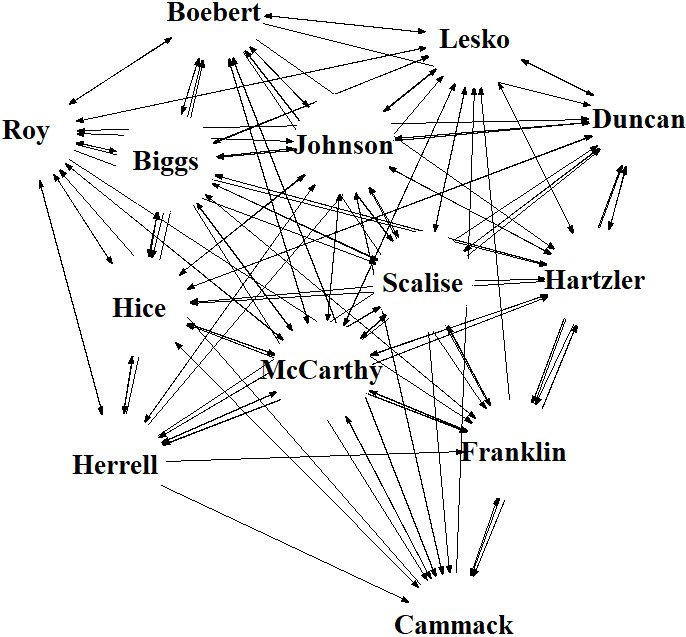}
				\label{fig:sub2}}
			\subfigure[Subgraph 3.]{
				\centering
				\includegraphics[scale=0.315]{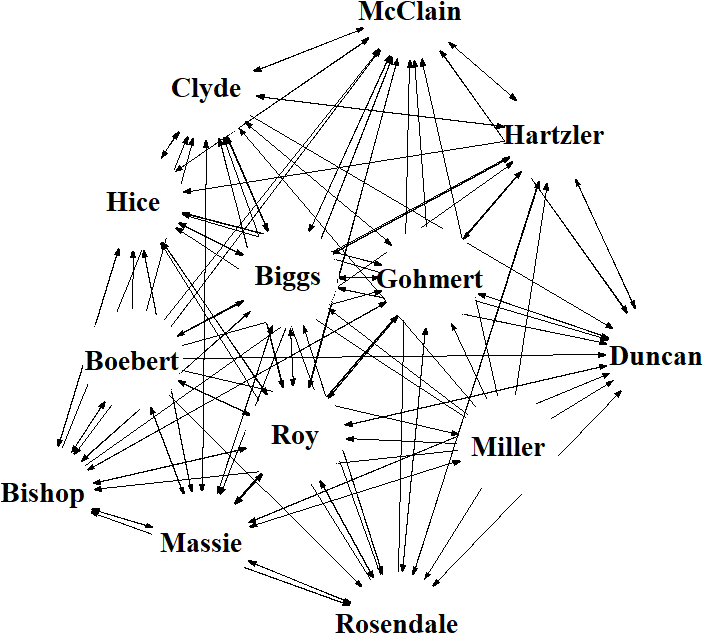}
				\label{fig:sub3}}
			
			\subfigure[Subgraph 4.]{
				\centering
				\includegraphics[scale=0.31]{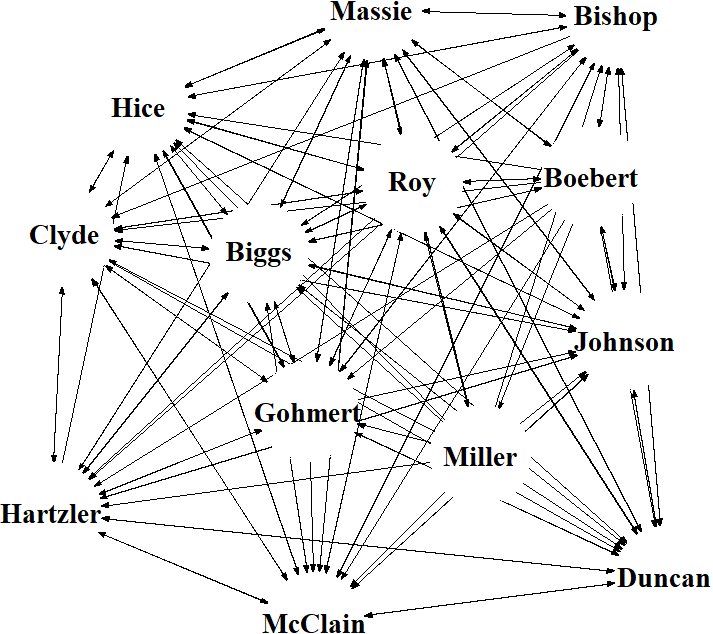}
				\label{fig:sub4}}
			\subfigure[Subgraph 5.]{
				\centering
				\includegraphics[scale=0.31]{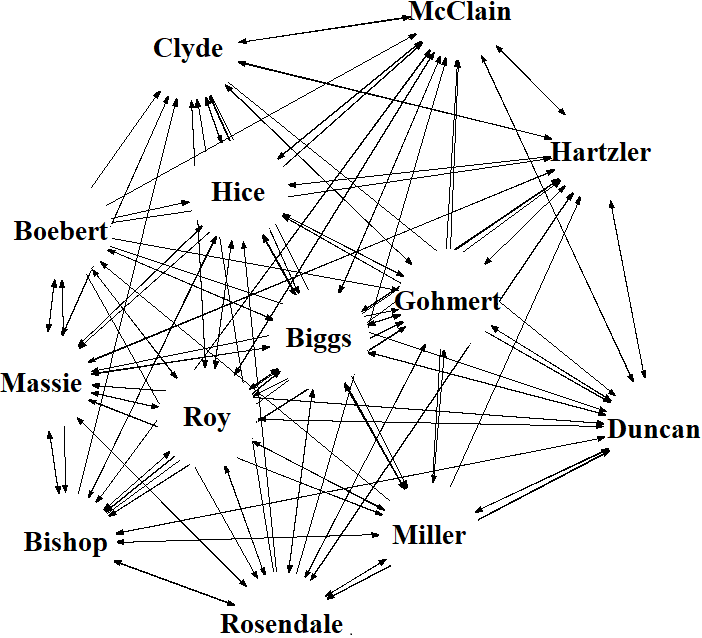}
				\label{fig:sub5}}
			\subfigure[Subgraph 6.]{
				\centering
				\includegraphics[scale=0.315]{sub3.png}
				\label{fig:sub6}}
			\caption{Maximal cliques in the network of the 117th Congress of the U. S. All the six subgroups are composed of members of the Republican Party.} 
			\label{fig:clanes}
		\end{center}
	\end{figure}

	The evaluation of triadic states shows that 27.97\% of the network's triads are closed, indicating frequent interactions among groups of three political actors. For instance, in the subgraphs in Figure \ref{fig:clanes}, we observe the triad formed by \textit{Roy}, \textit{Boebert}, and \textit{Biggs}--actors known for their opposition to immigration policies \citep[][]{H3807} and restrictions regarding firearm usage \citep[][]{congress}, among other issues of the Democrat agenda.
	Additionally, the proportion of mutual connections (reciprocity) is 46.15\%, indicating a strong tendency for bidirectional interactions between individuals. In Figure \ref{fig:clanes}, we can visualize several of these dyads. For example, we observe bidirectional interactions between Republicans \textit{Scalise} and \textit{McCarthy}, both of whom hold leadership roles in the House of Representatives.

	Additionally, we conduct a census of cliques \citep[cliques;][]{gjoka2014estimating} and determined that the largest subgraph \citep[maximal clique;][]{bouchitte2002listing} consists of 13 individuals. 
	We identify a total of six maximal cliques in the network (Figure \ref{fig:clanes}), all of which are composed of Republican Party members, most of them belonging to the House Freedom Caucus, a group of extremist legislators within the House of Representatives \citep[][]{green2019legislative}. 
	Some of them, such as \textit{Andy Biggs}, are also part of the Make America Great Again (MAGA) movement, a more radical faction within the Freedom Caucus loyal to Trump’s policies \citep[][]{eddington2018communicative}.
	These subgroups highlight interconnected individuals who share a high degree of affinity or cooperation, interact regularly with one another, and are crucial to maintaining the system’s cohesion \citep[][]{li2021review}. 
	We highlight \textit{Lauren Boebert}, \textit{Jeff Duncan}, \textit{Andy Biggs}, \textit{Jody Hice}, and \textit{Chip Roy} as common actors across the six subgroups. 
	The latter three had already been noted as authorities and important nodes based on their eigen centrality. Additionally, \textit{Boebert} is one of the network’s hubs, and \textit{Duncan} is one of the most prominent receivers.
	Figures \ref{fig:sub1} and \ref{fig:sub2} showcase other notable actors, such as \textit{McCarthy} and \textit{Franklin}. Furthermore, Figures \ref{fig:sub3}--\ref{fig:sub6} highlight \textit{Dan Bishop} as a key actor in the network’s connectivity.

	Using the global clustering coefficient, we find that of all possible connections between an actor's neighbors, the system exhibits approximately 26.95\%. 
	Particularly, the coefficient is 34.10\% for the Republican Party and 30.60\% for the Democrat Party. 
	Also, it stands at 40.00\% for the Senate and 29.80\% for the House of Representatives.
	This suggests that individuals within the system tend to form densely interconnected groups or communities and that the relationships emerging in the network do not follow a random pattern. Instead, they demonstrate an organized, cohesive structure \citep[][]{li2021review}.

	On the other hand, we examined the local clustering coefficient to identify actors who are highly interconnected within their neighborhoods \citep[][]{li2021review}. 
	For members of the network, this coefficient ranges from 0.000 to 0.600. 
	Nodes with a coefficient greater than 0.500 correspond to Republican Party representatives in the House, including \textit{Ben Cline} (0.591), \textit{Lance Gooden} (0.582), \textit{Paul Gosar} (0.581), \textit{Mary Miller} (0.565), \textit{Michael Cloud} (0.536), \textit{Virginia Foxx} (0.529), \textit{Brian Babin} (0.525), \textit{Larry Bucshon} (0.525), \textit{Ronny Jackson} (0.525), \textit{Diana Harshbarger} (0.524), \textit{Roger Williams} (0.511), and \textit{Steven Palazzo} (0.510).
	Among these, \textit{Cloud} and \textit{Miller} are part of the network's maximal clique (Figures \ref{fig:sub1}, \ref{fig:sub3}--\ref{fig:sub6}). Additionally, \textit{Cloud} serves as one of the system’s hubs.

	Members of Congress with a local clustering coefficient below 0.150 include \textit{Gregorio Kilili} (0.000), \textit{Dan Kildee} (0.089), \textit{Kay Granger} (0.095), \textit{Thomas Suozzi} (0.100), \textit{Ed Case} (0.107), \textit{Josh Gottheimer} (0.123), \textit{Tom O'Halleran} (0.133), \textit{Jim Himes} (0.140), \textit{Don Bacon} (0.143), \textit{Nancy Pelosi} (0.144), \textit{Marco Rubio} (0.144), and \textit{Kirsten Gillibrand} (0.146).
	Several of these actors are positioned on the periphery of the network (Figure \ref{fig:grado_fuerza}). With the exception of \textit{Granger}, \textit{Bacon}, and \textit{Rubio}, the rest are members of the Democrat Party.

	We examine the flow of information in the network using the assortativity, which is the tendency of individuals in the system to connect with others based on similarities in one or more node characteristics \citep{schwarzenbach2024extremists}. 
	For this analysis, we considered both node-level and structural variables of the network (Table \ref{tab:asortatividad}).
	We observed that both qualitative and quantitative node variables exhibit a positive assortativity coefficient. Notably, party affiliation and chamber are the node attributes with the highest assortativity coefficients. This suggests that network actors tend to connect with others who share similar characteristics in terms of party membership or chamber affiliation.

	Additionally, we observe that the structural characteristics of degree and betweenness centrality exhibit a negative assortativity coefficient. 
	This indicates that there are likely nodes with high scores connected to others with lower scores in these metrics. 
	All the other structural variables show a positive coefficient. 
	In particular, eigen centrality and hubs are the systematic properties with the highest assortativity. 
	This suggests that actors with high scores in these metrics tend to be connected with one another.

	\begin{table}[!htb]
		\scriptsize
		\centering
		\setlength{\tabcolsep}{2.5 pt}
		\begin{tabular}{llc}
			\hline
			\multicolumn{2}{c}{\textbf{Variable}}                             & \textbf{Asortativity} \\ \hline
			\multicolumn{1}{c}{\textit{Qualitative}} &                             &                        \\
			& \textit{Party}            & 0.647                  \\
			& \textit{Race}               & 0.149                  \\
			& \textit{Ethnicity}      & 0.156                  \\
			& \textit{Religion}           & 0.080                  \\
			& \textit{Sex}               & 0.143                  \\
			\textbf{}                                 & \textit{Chamber}             & 0.597                  \\
			& \textit{LGBTIQ+}  & 0.027                  \\
			& \textit{State}             & 0.096                  \\
			\textit{Quantitative}                    &                             &                        \\
			& \textit{Age}               & 0.101                  \\
			\textbf{}                                 & \textit{Time in Service} & 0.157                  \\
			\textit{Structural}                    &                             &                        \\
			& \textit{Out-degree}      & -0.025                 \\
			& \textit{In-degree}    & -0.043                 \\
			& \textit{Out-strength}   & 0.002                  \\
			& \textit{Closeness C.}        & 0.083                  \\
			& \textit{Betweenness C.}  & -0.028                 \\
			& \textit{Eigen C.}          & 0.371                  \\
			& \textit{Hubs}            & 0.414                  \\
			& \textit{Authorities}        & 0.190                  \\ \hline
		\end{tabular}
		\caption{Assortativity of the 117th Congress Twitter network of the United States, considering node-level (qualitative and quantitative) as well as structural variables.}
		\label{tab:asortatividad}
	\end{table}

	\subsection{Exponential Random Graph Model}

	Now, we fit a ERGM using structural predictors and nodal variables. 
	A global overview of the six models proposed here reveals that the systemic properties of the 117th Congress Twitter network of the United States play a crucial role in explaining how links emerge in the system (Tables \ref{tab:model_structure} and \ref{tab:model_structure1}). 
	We highlight that incorporating structural variables into the model reduces information criteria.
	Model 3 reduces the AIC and BIC by approximately 16\% compared to Model 1, which only considers the number of edges as a covariate. Similarly, Model 6 minimizes these criteria by about 25\% compared to Model 1, and by 3.5\% in contrast to Model 3. 
	Model 6 incorporates the structural characteristics that are statistically significant in Model 3, as well as the nodal variables of party and chamber.

	\begin{table}[!htb]
		\scriptsize
		\centering
		\setlength{\tabcolsep}{2pt} 
		\begin{tabular}{llcccccccc}
			\hline
			& \multicolumn{3}{c}{\textbf{Model 1}}                                               & \multicolumn{3}{c}{\textbf{Model 2}}                              & \multicolumn{3}{c}{\textbf{Model 3}}                                \\ \cline{2-10} 
			\textbf{Predictor}         & \multicolumn{1}{c}{\textbf{Estimate}} & \textbf{SD}          & \textbf{p-value}     & \textbf{Estimate}    & \textbf{SD}          & \textbf{p-value}     & \textbf{Estimate} & \textbf{SD}               & \textbf{p-value}     \\ \hline
			\textit{Edges}           & \multicolumn{1}{c}{-2.769}            & 0.009                & 0.000                & -3.354               & 0.012                & 0.000                & -5.375            & 0.060                     & 0.000                \\
			\textit{Reciprocity}      &                                       &                      &                      & 3.198                & 0.029                & 0.000                & 2.842             & 0.032                     & 0.000                \\
			\textit{In-degree}     &                                       &                      &                      &                      &                      &                      & 0.013             & 0.001                     & 0.000                \\
			\textit{Out-degree}      &                                       & \textbf{}            & \textbf{}            & \textbf{}            & \textbf{}            & \textbf{}            & 0.006             & 0.001                     & 0.000                \\
			\textit{Out-strength}     &                                       & \multicolumn{1}{l}{} & \multicolumn{1}{l}{} & \multicolumn{1}{l}{} & \multicolumn{1}{l}{} & \multicolumn{1}{l}{} & 0.181             & 0.097                     & 0.063                \\
			\textit{Closeness C.}       &                                       & \multicolumn{1}{l}{} & \multicolumn{1}{l}{} & \multicolumn{1}{l}{} & \multicolumn{1}{l}{} & \multicolumn{1}{l}{} & 0.988             & \multicolumn{1}{l}{0.068} & 0.000                \\
			\textit{Betweenness C.} &                                       & \textbf{}            & \textbf{}            & \textbf{}            & \textbf{}            & \textbf{}            & -7.948            & 0.680                     & 0.000                \\
			\textit{Eigen C.}         &                                       & \multicolumn{1}{l}{} & \multicolumn{1}{l}{} & \multicolumn{1}{l}{} & \multicolumn{1}{l}{} & \multicolumn{1}{l}{} & -0.775            & 0.455                     & 0.089                \\
			\textit{Hubs}           &                                       & \textbf{}            &                      &                      &                      &                      & 0.229             & 0.048                     & 0.001                \\
			\textit{Authorities}       &                                       & \multicolumn{1}{l}{} & \multicolumn{1}{l}{} & \multicolumn{1}{l}{} & \multicolumn{1}{l}{} & \multicolumn{1}{l}{} & 0.133             & 0.331                     & 0.687                \\
			&                                       & \textbf{}            & \textbf{}            &                      &                      &                      &                   &                           &                      \\
			AIC                        & \multicolumn{1}{c}{100991}            &                      &                      & 90960                &                      &                      & 84725             & \multicolumn{1}{c}{}      & \multicolumn{1}{l}{} \\
			BIC                        & \multicolumn{1}{c}{101001}            &                      &                      & 90980                & \textbf{}            & \textbf{}            & 84829             & \multicolumn{1}{l}{}      & \multicolumn{1}{l}{} \\ \hline
		\end{tabular}
		\caption{ERGMs 1, 2, and 3 using structural statistics as predictors for the Twitter network of the 117th U.S. Congress. SD stands for standard deviation.}
		\label{tab:model_structure}
	\end{table}

	The evaluated scenarios allow us to identify key factors in the configuration of the legislative network. 
	For instance, Model 2 highlights reciprocity as a relevant factor in explaining how links arise. 
	Meanwhile, Model 3 points out structural predictors that are not statistically significant, such as the out-degree, out-strength, eigen centrality, and authorities.
	In contrast, Model 4 shows that considering only nodal attributes does not improve the model's information criteria, compared to the Model 3 that incorporates only structural predictors. Furthermore, the latter is more parsimonious.

	On the other hand, Model 5 reveals that age and tenure are not statistically significant in explaining how links between political actors emerge. Previous studies on communication networks among members of the Brazilian Congress also emphasize that age does not play a prominent role in the formation of links within legislative networks \citep[][]{wojcik2019legislative}.
	In contrast, Model 6 shows that a combination of structural characteristics and nodal attributes, such as party affiliation and legislative chamber, improves model performance compared to the other scenarios. 
	The prominence of party and chamber aligns with the results of the assortativity analysis (Table \ref{tab:asortatividad}), and their inclusion allows us to analyze homophilic effects in the configuration of the system.
	Even though we explored the influence of various nodal attributes together with structural characteristics on the overall configuration of the network, none of the alternative scenarios yielded better results than Model 6. This suggests that the combination of both structural predictors and key nodal attributes like party affiliation and legislative chamber in Model 6 offers the most comprehensive explanation for how connections form within the network.

	\begin{table}[!htb]
		\scriptsize
		\centering
		\setlength{\tabcolsep}{0.5 pt} 
		\begin{tabular}{llccccccccc}
			\hline
			\multicolumn{2}{l}{}                                   & \multicolumn{3}{c}{\textbf{Model 4}}              & \multicolumn{3}{c}{\textbf{Model 5}}              & \multicolumn{3}{c}{\textbf{Model 6}}              \\ \cline{3-11} 
			\multicolumn{2}{c}{\textbf{Predictor}}                 & \textbf{Estimate} & \textbf{SD} & \textbf{p-value} & \textbf{Estimate} & \textbf{SD} & \textbf{p-value} & \textbf{Estimate} & \textbf{SD} & \textbf{o value} \\ \hline
			\textit{Edges}           & \textit{}                 & -5.130            & 0.098       & 0.000            & -5.215            & 0.087       & 0.000            & -7.290            & 0.070       & 0.000            \\
			\textit{Reciprocity}      & \textit{}                 &                   &             &                  & 2.846             & 0.031       & 0.000            & 2.251             & 0.034       & 0.000            \\
			\textit{In-degree}     & \textit{}                 &                   &             &                  & 0.012             & 0.001       & 0.000            & 0.013             & 0.001       & 0.000            \\
			\textit{Out-degree}      & \textit{}                 &                   &             &                  & 0.007             & 0.001       & 0.000            & 0.009             & 0.001       & 0.000            \\
			\textit{Closeness C.}       & \textit{}                 &                   &             &                  & 0.969             & 0.061       & 0.000            & 1.039             & 0.069       & 0.000            \\
			\textit{Betweenness C.} & \textit{}                 &                   &             &                  & -9.132            & 0.646       & 0.000            & -9.709            & 0.716       & 0.000            \\
			\textit{Hubs}           & \textit{}                 &                   &             &                  & 0.201             & 0.047       & 0.000            & 0.275             & 0.063       & 0.000            \\
			\textit{Age}              & \textit{}                 & -0.003            & 0.001       & 0.000            & -0.001            & 0.001       & 0.517            &                   &             &                  \\
			\textit{Time in Service}   & \textit{}                 & 0.010             & 0.001       & 0.000            & -0.000            & 0.001       & 0.230            &                   &             &                  \\
			\textit{Party}           & \textit{}                 &                   &             &                  &                   &             &                  &                   &             &                  \\
			\textit{}                  & \textit{Democrat}        & 1.594             & 0.027       & 0.000            &                   &             &                  & 1.499             & 0.027       & 0.000            \\
			\textit{}                  & \textit{Republican}      & 1.960             & 0.028       & 0.000            &                   &             &                  & 1.555             & 0.027       & 0.000            \\
			\textit{}                  & \textit{Independent}    & -Inf              & 0.000       & 0.000            &                   &             &                  &                   &             &                  \\
			\textit{Race}              & \textit{}                 &                   &             &                  &                   &             &                  &                   &             &                  \\
			\textit{}                  & \textit{White}           & 0.086             & 0.025       & 0.001            &                   &             &                  &                   &             &                  \\
			\textit{}                  & \textit{Black}            & 0.549             & 0.053       & 0.000            &                   &             &                  &                   &             &                  \\
			\textit{}                  & \textit{Native American} & 0.045             & 1.064       & 0.966            &                   &             &                  &                   &             &                  \\
			\textit{}                  & \textit{Asiatic}         & 1.262             & 0.180       & 0.000            &                   &             &                  &                   &             &                  \\
			\textit{}                  & \textit{Other}             & 0.581             & 0.310       & 0.061            &                   &             &                  &                   &             &                  \\
			\textit{Ethnicity}     & \textit{}                 &                   &             &                  &                   &             &                  &                   &             &                  \\
			\textit{}                  & \textit{Hispanic}          & 0.810             & 0.319       & 0.011            &                   &             &                  &                   &             &                  \\
			\textit{}                  & \textit{Not Hispanic}       & 0.154             & 0.033       & 0.000            &                   &             &                  &                   &             &                  \\
			\textit{Religion}          & \textit{}                 &                   &             &                  &                   &             &                  &                   &             &                  \\
			\textit{}                  & \textit{Christian}        & -0.148            & 0.024       & 0.000            &                   &             &                  &                   &             &                  \\
			\textit{}                  & \textit{Other}             & 0.255             & 0.0056      & 0.000            &                   &             &                  &                   &             &                  \\
			\textit{Sexo}              & \textit{}                 &                   &             &                  &                   &             &                  &                   &             &                  \\
			\textit{}                  & \textit{Female}         & 0.508             & 0.030       & 0.000            &                   &             &                  &                   &             &                  \\
			\textit{}                  & \textit{Male}        & 0.004             & 0.021       & 0.840            &                   &             &                  &                   &             &                  \\
			\textit{Chamber}            & \textit{}                 &                   &             &                  &                   &             &                  &                   &             &                  \\
			\textit{}                  & \textit{Upper}             & 2.416             & 0.039       & 0.000            &                   &             &                  & 1.991             & 0.036       & 0.000            \\
			\textit{}                  & \textit{Lower}             & 1.166             & 0.028       & 0.000            &                   &             &                  & 0.908             & 0.027       & 0.000            \\
			\textit{LGBTIQ+}           & \textit{}                 &                   &             &                  &                   &             &                  &                   &             &                  \\
			\textit{}                  & \textit{Yes}               & 0.434             & 0.409       & 0.288            &                   &             &                  &                   &             &                  \\
			\textit{}                  & \textit{No}               & 0.159             & 0.055       & 0.004            &                   &             &                  &                   &             &                  \\
			&                           &                   &             &                  &                   &             &                  &                   &             &                  \\
			AIC                        &                           & 88643             &             &                  & 84782             &             &                  & 75698             &             &                  \\
			BIC                        &                           & 88849             &             &                  & 84875             &             &                  & 75811             &             &                  \\ \hline
		\end{tabular}
		\caption{ERGMs 4, 5, and 6 using structural statistics and node attributes as predictors for the Twitter network of the 117th U.S. Congress. SD stands for standard deviation.}
		\label{tab:model_structure1}
	\end{table}

	We implement Model 6 with a party reassignment for the four political actors who serve as independent congress members. 
	The limited information available for this category does not guarantee the estimability of coefficients (see Model 4) or information criteria in some cases. 
	Therefore, we opt for recategorization to avoid losing information and to enable inferences about the model’s parameters.
	As a result, legislators \textit{Angus King}, \textit{Bernie Sanders}, and \textit{Kyrsten Sinema} are assigned to the Democrat Party, while \textit{Jenniffer González} is assigned to the Republican Party. 
	These assignments are made based on political affinity and career trajectory.
	Nevertheless, we run the model considering all 16 possible reassignments (not shown here), and in every scenario, we obtained results analogous to those of Model 6 (Table \ref{tab:model_structure1}). 
	Therefore, we conclude that the proposed recategorization does not introduce any bias.

	\begin{table}[!htb]
		\scriptsize
		\centering
		\setlength{\tabcolsep}{2 pt} 
		\begin{tabular}{llccc}
			\hline
			\textit{}                  & \textit{}            & \textbf{Parameter $\theta$} & \textbf{exp$(\theta)$} & \textbf{expit$(\theta)$} \\ \hline
			\textit{Edges}           & \textit{}            & -7.290                      & 0.001        & 0.001          \\
			\textit{Reciprocity}      & \textit{}            & 2.251                       & 9.497        & 0.905          \\
			\textit{In-degree}     & \textit{}            & 0.013                       & 1.013        & 0.503          \\
			\textit{Out-degree}      & \textit{}            & 0.009                       & 1.009        & 0.502          \\
			\textit{Closeness C.}       & \textit{}            & 1.039                       & 2.828        & 0.739          \\
			\textit{Betweenness C.} & \textit{}            & -9.709                      & 0.001        & 0.001          \\
			\textit{Hubs}           & \textit{}            & 0.275                       & 1.317        & 0.568          \\
			\textit{Party}           & \textit{Democrat}   & 1.499                       & 4.477        & 0.817          \\
			\textit{}                  & \textit{Republican} & 1.555                       & 4.733        & 0.826          \\
			\textit{Chamber}            & \textit{Upper}        & 1.991                       & 7.326        & 0.880          \\
			\textit{}                  & \textit{Lower}        & 0.908                       & 2.480        & 0.713          \\ \hline
		\end{tabular}
		\caption{Probabilities associated with the parameters of Model 6.}
		\label{tab:prop}
	\end{table}

	All predictors in Model 6 are statistically significant using a 5\% significance level. 
	The signs of the estimated parameters for party and chamber align with the expected direction. 
	Both homophilic effects are positive, meaning that it is highly likely for congress members from the same party or chamber to interact with one another.
	Specifically, keeping the values of other statistics constant, we find that the probability (in logit scale) of observing an interaction between two Democrats is 1.499. This means that being a member of the Democrat Party increases the odds of interaction by a factor of 4.477, which corresponds to an 81.7\% increase in the probability of observing a particular interaction (Table \ref{tab:prop}). 
	The coefficient for the Republican Party is interpreted similarly, showing that the contribution of both parties to the probability of forming interactions within the network is comparable.
	In terms of legislative chamber, we see that belonging to the Senate increases the odds of interaction by a factor of 7.326, which corresponds to an 88.0\% increase in the probability of observing a specific interaction. 
	Meanwhile, being a member of the lower chamber increases the odds of interaction by a factor of 2.480, with a probability increase of 71.3\%. 
	Previous studies have noted that physical proximity enhances the likelihood of forming links within a legislative network \citep[][]{wojcik2019legislative}.

	Reciprocity, in- and out-degree, closeness centrality, and hubs also have a positive effect on the probability of observing a particular interaction within the network. 
	Notably, we observe that the Twitter network of the 117th U.S. Congress exhibits a high degree of reciprocity. 
	After homophilic effects, reciprocity is the most significant factor impacting the probability of interactions within the network (Table \ref{tab:prop}). 
	Thus, if node $i$ is linked to node $j$, it is highly likely that $j$ will also be linked to $i$. 
	On the other hand, we find that the number of edges and betweenness centrality have a negative effect on the probability of interaction. 
	However, their contribution to this probability on a real scale is almost negligible.

	\begin{figure}[!htb]
		\centering
		\includegraphics[scale=0.65]{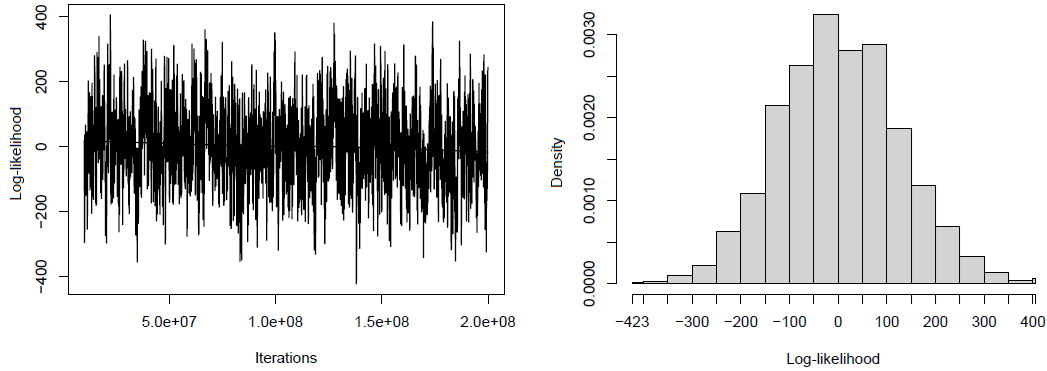}
		\caption{Markov chain and log-likelihood distribution for the number of edges in Model 6. The convergence diagnostics for the other predictors in the model exhibit similar behavior.}
		\label{fig:convergencia}
	\end{figure}
	
	The convergence diagnostics for Model 6, as well as for the other models, indicate no issues with stationarity in the chains and reveal an approximately normal distribution for the log-likelihood of all predictors. 
	In Figure \ref{fig:convergencia}, we present the Markov chain and the log-likelihood distribution for the number of edges. This behavior is analogous for the other covariates in the model.
	Additionally, the analysis of variance yields a p-value of less than 5\%. This indicates that Model 6 demonstrates superior predictive ability and goodness of fit compared to the model that suggests links emerge in a completely random manner.

	\subsection{Stochastic Block Model}

\begin{figure}[!htb]
		\centering
		\includegraphics[width=0.55\linewidth]{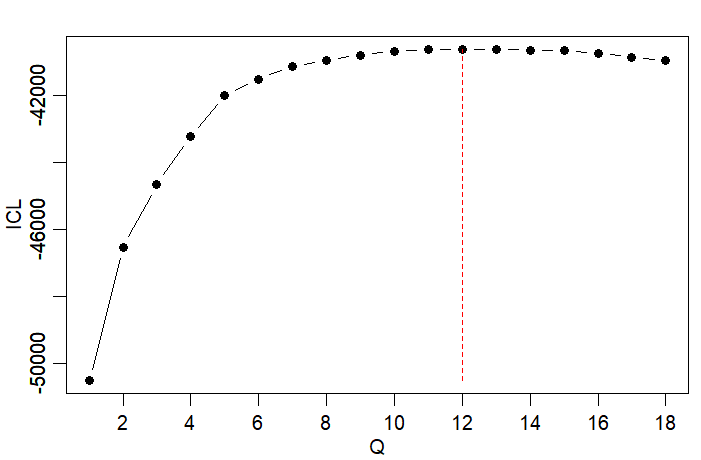}
		\caption{Integrated conditional likelihood. The red line indicates the optimal number of communities.}
		\label{fig:icl}
	\end{figure}

\begin{figure}[!htb]
		\scriptsize
		\begin{center}
			\begin{minipage}{0.5\textwidth}
				\centering
				\includegraphics[scale=0.45]{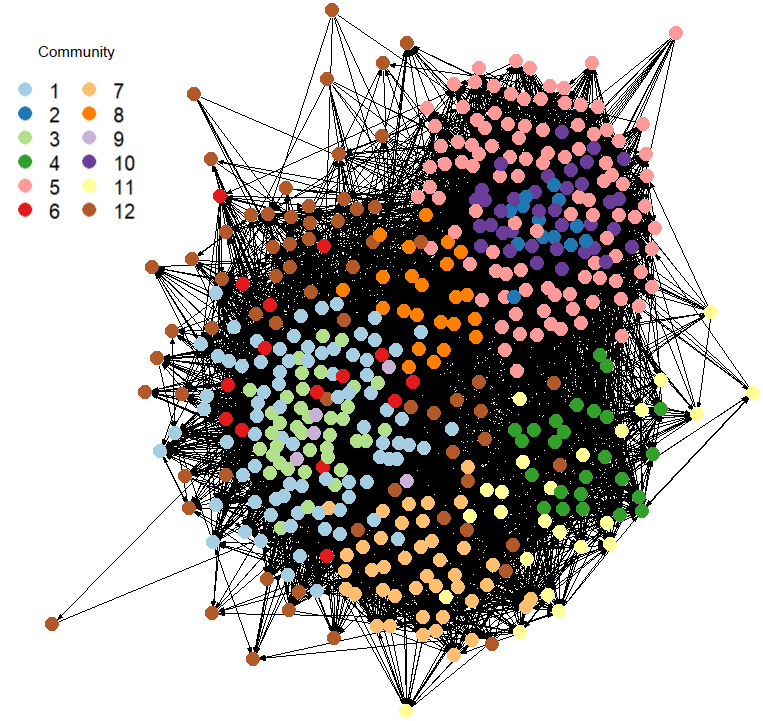}
				\caption{Partition induced by the SBM.}
				\label{fig:sbm}
			\end{minipage}\hfill   
			\begin{minipage}{0.45\textwidth}
				\centering
				\begin{tabular}{ccc}
					\\ \\
					\hline
					\textbf{Community} & \textbf{Size} & \textbf{Proportion} \\
					\hline
					\\
					1  & 82 & 17.26 \\
					2  & 13 &  2.74\\
					3  & 38 &  8.00\\
					4  & 25 &  5.26\\
					5  & 105 & 22.11\\
					6  & 15 &  3.16\\
					7  & 41 &  8.63\\
					8  & 23 &  4.84\\
					9  & 5 &  1.05\\
					10  & 36 &  7.58\\
					11  & 25 &  5.26\\
					12  & 67 &  14.11\\
					\\
					\hline
				\end{tabular}
				\captionof{table}{Distribution of the community structure.}
				\label{tab:community}
			\end{minipage}
		\end{center}
	\end{figure}

	Now, we implement a SBM in order to analyze the community structure in the Twitter network of the 117th U.S. Congress. To achieve this, we begin by optimizing the integrated conditional likelihood \citep[ICL; e.g.,][]{kolaczyk2014statistical} to determine the optimal number of communities. We obtain an optimal number of 12 partitions (Figure \ref{fig:icl}). We identify that communities 1, 5, and 12 include more than 50\% of the political actors (Table \ref{tab:community}), and we observe that groups 6, 11, and 12 are more dispersed within the system compared to the other groups (Figure \ref{fig:sbm}).
	
	Based on the partition, we observe that communities 2, 5, and 10 are positioned on the right side of the network, where there is a higher concentration of Republican Party members (Figure \ref{fig:sbm}). 
	These groups exhibit a concentric pattern and consist entirely of members of the House of Representatives who are not affiliated with the LGBTIQ+ community.
	Most of their members are of Christian affiliation (92.31\%, 97.14\%, and 94.44\%, respectively, in order), non-Hispanic (100\% in groups 2 and 10, and 96.19\% in group 5), and at least 75\% are white men.
	Communities 4 and 11, located in the lower-right part of the network, show a similar distribution to communities 2, 5, and 10 in terms of their nodal variables. However, all members of community 4 are part of the Senate. In community 11, Republicans are the majority (68.00\%), with 96.00\% being from the upper chamber.

	On the left side of the network, we observe communities 1, 3, and 9, which also display a roughly concentric arrangement (Figure \ref{fig:sbm}). 
	This portion shows a high concentration of Democrat supporters. 
	However, community 3 includes a Republican member, \textit{Katherine Castor} from Florida.
	These three communities are predominantly composed of members of the House of Representatives, with only community 9 including a Senate member, Democrat \textit{Chuck Schumer}. 
	In communities 6 and 12, Democrats also predominate (86.67\% and 90.04\%, respectively) within the lower chamber (93.33\% and 100\%, respectively). 
	These communities feature non-Hispanic legislators with at least 86.00\% representation and Christian affiliation with a minimum participation of 53.00\%.

	we observe In communities 1, 3, and 9 a more equitable gender distribution compared to what we see in groups 2, 5, and 10. 
	At least 40.00\% of the legislators are female, with community 1 having a female representation of 53.66\%. 
	Additionally, there is representation of LGBTIQ+ members in communities 1 and 9, at 2.44\% and 20.00\%, respectively.
	Compared to communities 2, 5, and 10, the proportion of white legislators decreases, with a maximum of 60.00\% in communities 1, 3, and 9. 
	Community 7, located in the lower part of the network, has a distribution similar to these communities in terms of nodal variables, except that all its members are from the Senate and there is a lower female representation (34.15\%).
	Furthermore, community 8, located between groups 1 and 5, consists exclusively of members from the lower chamber and has no representatives from the LGBTIQ+ community. It includes Republican supporters (65.22\%), Democrats (30.43\%), and Independents (4.35\%). This community also features a predominance of white men (60.87\%) and Christian affiliation (82.61\%).

	In Table \ref{tab:media_comunidad}, we highlight the dominant party and chamber for each community. 
	75.00\% percent of the communities are dominated by members of the House of Representatives, while party dominance is quite balanced, with each party prevailing in 6 of the 12 communities. 
	Groups 8 and 11 have the highest number of members from both parties.
	We also present the mean and coefficient of variation (CV) of the structural variables for each community. 
	We observe high variability in betweenness centrality across all groups. Additionally, we identify communities 2 and 9 as those with the highest number of metrics exhibiting high average values.

	\begin{table}[!htb]
		\scriptsize
		\centering
		\setlength{\tabcolsep}{2.5 pt}
		\begin{tabular}{llcccccccccccc}
			\hline
			&       & \multicolumn{12}{c}{\textbf{Community}}                                                                                                                                         \\ \cline{3-14} 
			\textbf{}                                    &       & \textbf{1} & \textbf{2}                    & \textbf{3} & \textbf{4} & \textbf{5} & \textbf{6} & \textbf{7} & \textbf{8} & \textbf{9} & \textbf{10} & \textbf{11} & \textbf{12} \\ \hline \\
			\textit{Party \%}                         & \textit{} & \begin{tabular}[c]{@{}c@{}}D\\ 98\end{tabular}     & \begin{tabular}[c]{@{}c@{}}R\\ 100\end{tabular}    & \begin{tabular}[c]{@{}c@{}}D \\ 97\end{tabular}    & \begin{tabular}[c]{@{}c@{}}R\\ 100\end{tabular}    & \begin{tabular}[c]{@{}c@{}}R\\ 99\end{tabular}     & \begin{tabular}[c]{@{}c@{}}D\\ 87\end{tabular}    & \begin{tabular}[c]{@{}c@{}}D \\ 93\end{tabular}    & \begin{tabular}[c]{@{}c@{}}R\\ 65\end{tabular}     & \begin{tabular}[c]{@{}c@{}}D\\ 100\end{tabular}   & \begin{tabular}[c]{@{}c@{}}R\\ 100\end{tabular}    & \begin{tabular}[c]{@{}c@{}}R\\ 68\end{tabular}    & \begin{tabular}[c]{@{}c@{}}D\\ 91\end{tabular}     \\ \\
			\textit{Chamber \%}                          & \textit{} & \begin{tabular}[c]{@{}c@{}}Lower\\ 100\end{tabular} & \begin{tabular}[c]{@{}c@{}}Lower\\ 100\end{tabular} & \begin{tabular}[c]{@{}c@{}}Lower\\ 100\end{tabular} & \begin{tabular}[c]{@{}c@{}}Upper\\ 100\end{tabular} & \begin{tabular}[c]{@{}c@{}}Lower\\ 100\end{tabular} & \begin{tabular}[c]{@{}c@{}}Lower\\ 93\end{tabular} & \begin{tabular}[c]{@{}c@{}}Upper\\ 100\end{tabular} & \begin{tabular}[c]{@{}c@{}}Lower\\ 100\end{tabular} & \begin{tabular}[c]{@{}c@{}}Lower\\ 80\end{tabular} & \begin{tabular}[c]{@{}c@{}}Lower\\ 100\end{tabular} & \begin{tabular}[c]{@{}c@{}}Upper\\ 96\end{tabular} & \begin{tabular}[c]{@{}c@{}}Lower\\ 100\end{tabular} \\ \\
			
			& Media & 0.233      & 0.723 & 0.469      & 0.219      & 0.134      & 0.059      & 0.211      & 0.248      & 0.304      & 0.419       & 0.067       & 0.078       \\
			\multirow{-2}{*}{\textit{In-degree}}     & CV \%    & 33         & 28                            & 34         & 48         & 49         & 44         & 42         & 37         & 70         & 28          & 57          & 59          \\
			& Media & 0.112      & 0.303                         & 0.204      & 0.144      & 0.105      & 0.165      & 0.151      & 0.126      & 0.557      & 0.164       & 0.085       & 0.051       \\
			\multirow{-2}{*}{\textit{Out-degree}}      & CV \%    & 41         & 52                            & 32         & 35         & 41         & 29         & 35         & 36         & 45         & 32          & 47          & 50          \\
			& Media & 0.132      & 0.333                         & 0.183      & 0.137      & 0.192      & 0.267      & 0.178      & 0.187      & 0.690      & 0.206       & 0.142       & 0.078       \\
			\multirow{-2}{*}{\textit{Out-strength}}     & CV \%    & 67         & 87                            & 54         & 69         & 62         & 51         & 44         & 88         & 40         & 47          & 57          & 64          \\
			& Media & 0.336      & 0.466                         & 0.479      & 0.493      & 0.201      & 0.247      & 0.439      & 0.323      & 0.466      & 0.318       & 0.265       & 0.204       \\
			\multirow{-2}{*}{\textit{Closeness C.}}       & CV \%    & 35         & 38                            & 27         & 40         & 61         & 62         & 33         & 55         & 51         & 45          & 45          & 63          \\
			& Media & 0.025      & 0.273                         & 0.151      & 0.098      & 0.007      & 0.007      & 0.059      & 0.071      & 0.145      & 0.050       & 0.003       & 0.003       \\
			\multirow{-2}{*}{\textit{Betweenness C.}} & CV \%   & 204        & 101                           & 143        & 172        & 399        & 223        & 161        & 161        & 155        & 133         & 318         & 249         \\
			& Media & 0.024      & 0.458                         & 0.043      & 0.045      & 0.145      & 0.025      & 0.017      & 0.096      & 0.076      & 0.265       & 0.019       & 0.016       \\
			\multirow{-2}{*}{\textit{Eigen C.}}         & CV \%    & 52         & 47                            & 34         & 75         & 53         & 42         & 68         & 62         & 76         & 27          & 67          & 94          \\
			& Media & 0.023      & 0.293                         & 0.032      & 0.052      & 0.219      & 0.045      & 0.021      & 0.116      & 0.136      & 0.259       & 0.034       & 0.023       \\
			\multirow{-2}{*}{\textit{Hubs}}           & CV \%    & 72         & 74                            & 62         & 143        & 76         & 50         & 61         & 97         & 82         & 49          & 71          & 139         \\
			& Media & 0.020      & 0.449                         & 0.040      & 0.032      & 0.056      & 0.006      & 0.012      & 0.052      & 0.016      & 0.206       & 0.005       & 0.008       \\
			\multirow{-2}{*}{\textit{Authorities}}       & CV \%    & 88         & 53                            & 51         & 84         & 72         & 106        & 150        & 56         & 72         & 37          & 111         & 92          \\ \\ \hline
		\end{tabular}
		\caption{Summary of nodal variables and structural characteristics by community. For each group, we present the dominant party and chamber: Democrats (D) and Republicans (R). We display the mean and coefficient of variation (CV) of the network's structural variables for each community. We normalize the values of each metric to a range of 0 to 1.}
		\label{tab:media_comunidad}
	\end{table}

	Community 2 shows high average values across all the evaluated metrics. 
	It is composed of \textit{Jim Banks}, \textit{Andy Biggs}, \textit{Byron Donalds}, \textit{Jeff Duncan}, \textit{Tom Emmer}, \textit{Scott Franklin}, \textit{Mike Johnson}, \textit{Kevin McCarthy}, \textit{Mary Miller}, \textit{John Rose}, \textit{Chip Roy}, \textit{Steve Scalise}, and \textit{Elise Stefanik}. 
	Some of these members were previously identified as key actors in the system's configuration (Section \ref{sec:actores}). Members of this group represent 61.54\% of the political figures in the maximal clan (Figure \ref{fig:clanes}).
	In contrast, community 9 exhibits high average values in six of the eight metrics considered and shows average values for eigenvector centrality and authority. It includes \textit{Chuck Schumer}, \textit{Rosa DeLauro}, \textit{Nancy Pelosi}, \textit{Bobby Rush}, and \textit{Mark Takano}. 
	All these members hold progressive and liberal positions on economic, health, education, civil rights, and social policy issues. 
	For instance, \textit{Takano} is co-chair of the \textit{Congressional LGBTQ+ Equality Caucus}, which works for the rights of the LGBTQ+ community \citep[][]{bergersen2018intersectionality}. 
	Meanwhile, \textit{Rush} is a member of the \textit{Afterschool Caucuses} and advocates for policies benefiting African American, low-income, or vulnerable communities \citep[e.g.,][]{marcaurelle1999payday, davies2011legislative}.

	Groups 3 and 10 are notable for having high average values across many metrics. Community 3 includes actors who excel as receivers, senders, and bridges, playing crucial roles in connecting and interacting with other key individuals in the network. Meanwhile, community 10 features a mix of senders, receivers, and centers of influence, with its members being highly referenced and well-connected, further enhancing their impact within the network.

	Our findings reveal that the communities which include the predominant actors in the network based on systematic properties are 2, 3, 9, and 10.
	Communities with actors of medium relevance are 4, 5, 6, and 8. Communities 1 and 7 consist of actors with medium-low relevance, while communities 11 and 12 concentrate less relevant actors according to the network’s systematic properties. 
	For example, community 12 includes delegate \textit{Gregorio Kilili Camacho}, who was previously noted for lacking an influential role in the system (Section \ref{sec:actores})

	\begin{figure}[!htb]
		\scriptsize
		\centering
		\includegraphics[width = 0.67 \linewidth]{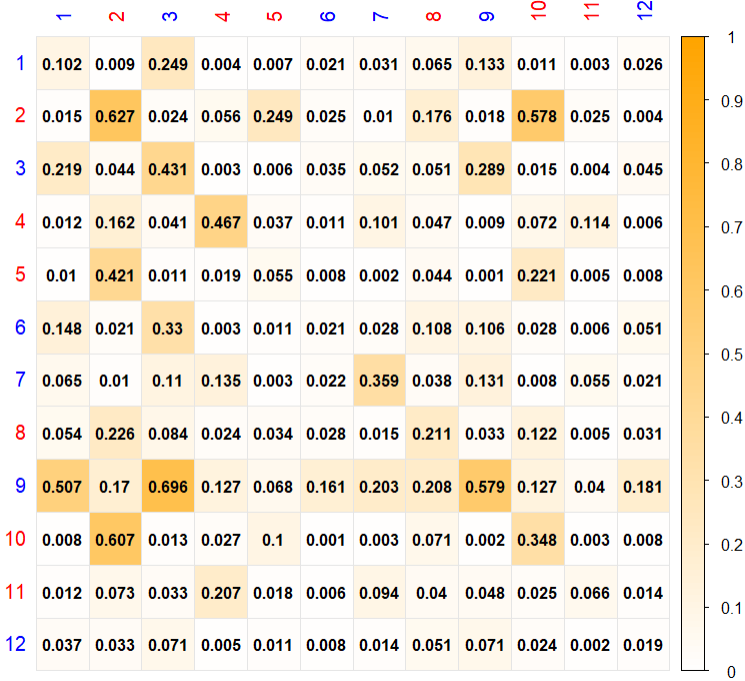}
		\caption{Interaction probability matrix between Communities. Communities with a predominance of Republicans are labeled in blue, while those with a higher presence of Democrats are labeled in red.}
		\label{fig:prop_inter}
	\end{figure}

	Additionally, we estimate the probabilities of intra- and inter-community interactions (Figure \ref{fig:prop_inter}). 
	We observe that actors in groups 2 and 9 have a high probability of intra-community connections, at 62.7\% and 57.9\%, respectively. 
	Communities 3, 4, 7, and 10 have intra-community interaction probabilities of 43.1\%, 46.7\%, 35.9\%, and 38.4\%, respectively. 
	Specifically, groups 2, 3, 4, and 5 show higher probabilities of intra-group interactions compared to their inter-group interaction probabilities. 
	Conversely, we identify the opposite pattern for the remaining groups.

	There are communities where the probability of interaction between them is nearly zero. 
	For example, it is highly unlikely that members of community 5 will connect with those in community 7, or that members of community 4 will connect with those in community 12. 
	Group 5 is led by Republican House members, while community 7 is composed of Senate Democrats (Table \ref{tab:media_comunidad}). 
	Communities 4 and 12 exhibit a similar pattern in terms of chamber and party dominance. Additionally, interactions between members of different parties or chambers are generally unlikely. 
	Communities 4 and 6, and 2 and 12, demonstrate this pattern. 
	However, the latter two groups differ only by party, as their members are predominantly from the lower chamber.

	We also observe interactions that are more likely to occur in one direction. For example, communities 1 and 3 have a low probability of connecting with communities 4 and 11. However, there is a higher likelihood that groups 4 and 11 will connect with groups 1 and 3 (Figure \ref{fig:prop_inter}). 
	The latter are led by House Democrats, while groups 4 and 11 are predominantly Senate Republicans (Table \ref{tab:media_comunidad}). 
	Additionally, it is more likely that the Republican minority in the Senate will contact the Democrat majority in the House, rather than the connection occurring in the opposite direction.

	The probabilities of interaction between communities vary depending on the direction of the link, even when they involve members from the same party or chamber. 
	For instance, we find that it is more likely for an actor from group 9 to interact with one from group 3 (70.00\%) than for the interaction to occur in the opposite direction (29.00\%). 
	These interaction probabilities align with the roles predominant in each group based on the average values of the evaluated metrics. 
	Specifically, actors in group 9 predominantly play the roles of senders and influencers (high out-degree and strength), while group 3 actors are more likely to be receivers of information (high in-degree) within the Democrat Party. 
	This explains why interactions are more likely to flow from group 9 to group 3.

	Additionally, we observe that communities 2 and 10 have similar probabilities of interaction in both directions (Figure \ref{fig:prop_inter}). 
	However, when comparing the average values of structural variables for these two groups, it is clear that actors in community 2 play crucial roles within the Republican Party as receivers, influencers, senders, and intermediaries, and are well-connected with other important nodes. 
	In contrast, actors in community 10 exhibit moderate roles in the last three characteristics (Table \ref{tab:media_comunidad}).

	Finally, we compared the natural clustering derived from nodal variables with the clustering induced by the SBM (Table \ref{tab:contraste}). 
	The evaluation criteria indicate that the SBM induced partition captures the characteristics of party affiliation and chamber. 
	However, there is no significant alignment between the clusters for other variables.

	\begin{table}[!htb]
		\scriptsize
		\centering
		\setlength{\tabcolsep}{2 pt} 
		\begin{tabular}{lccccccc}
			\cline{2-8}
			& \multicolumn{7}{c}{\textbf{Qualitative nodal variables}}                                                                        \\ \hline
			\multicolumn{1}{c}{\textbf{Metric}} & \textit{Party} & \textit{Race} & \textit{Ehnicity} & \textit{Religion} & \textit{Sex} & \textit{Chamber} & \textit{LGBTIQ+} \\ \hline
			\textit{Rand}                        & 0.615            & 0.448         & 0.237                  & 0.316             & 0.455         & 0.438           & 0.156            \\
			\textit{Adjusted.rand}               & 0.220            & 0.020         & 0.001                  & 0.016             & 0.026         & 0.122           & 0.002            \\
			\textit{nmi}                         & 0.359            & 0.086         & 0.018                  & 0.039             & 0.039         & 0.345           & 0.014         \\
			\hline  
		\end{tabular}
		\caption{\textit{Comparison between the natural clustering and SBM induced partition.}}
		\label{tab:contraste}
	\end{table}

	\section{Discussion} \label{sec:discussion}

	In this study, we used Twitter interaction data among members of the 117th U.S. Congress to evaluate the visibility of political leaders and explore how systemic properties and nodal attributes influence network link formation in online legislative networks. 
	This research addresses a gap in the current literature, which primarily focuses on explaining political connections from a partisan perspective, without delving deeply into other institutional traits, individual attributes, and systemic properties that determine leadership positioning and relationships in online legislative networks. 
	Understanding how politicians connect, considering factors beyond partisanship, enhances our comprehension of political participation phenomena outside traditional legislative arenas \citep[][]{cook2016american, hemphill2021drives}.

	Our analysis identified actors with leadership roles as emitters, receivers, and influencers. These figures hold official positions in dominant parties and are well-known and experienced in Congress. For instance, \textit{Kevin McCarthy}, Republican minority leader, stands out with high scores in the out-degree and out-strength metrics. \textit{Nancy Pelosi}, Democrat and Speaker of the House, is not a notable emitter but excels as a receiver and influencer. Other prominent emitters and influencers include Democrat \textit{Bobby Rush} and Republican whip \textit{Steve Scalise}, who have served in Congress for 30 and 16 years, respectively.

	Centrality metrics such as closeness and betweenness reveal that both parties have significant actors who facilitate communication and information transfer within the system. Metrics of eigen centrality, authorities, and hubs indicate that the Republican minority in the House of Representatives holds significant leadership, being the most important figures based on their direct connections and social presence. Additionally, politicians with less influence in the system are often associated with lower-profile Congressional roles or lesser-known individuals, like independent members and delegates such as \textit{Gregorio Kilili Camacho}.

	Our findings highlight leadership associated with institutional hierarchy, party affiliation, and chamber, but no distinct leadership pattern emerges from other individual attributes evaluated. Although we did not test party-specific hypotheses, our results show that Twitter connections among legislators continue to reflect the partisan trends reported in previous studies \citep[][]{cook2016american, borge2017opinion, praet2021patterns}. Similar to other digital studies, we argue that social networks on platforms like Twitter reinforce rather than weaken the leadership of dominant political figures \citep[][]{grant2010digital}, though these leadership roles can vary.

	A limitation of our study is the lack of differentiation between types of network connections, such as those creating information (tweets), disseminating information (retweets), or fostering conversations through mentions (@) and hashtags (\#) \citep[see][]{borge2017opinion, del2018echo, praet2021patterns}. This differentiation would allow for a more specific characterization of leadership roles by considering the intent of connections. Additionally, we did not analyze tweet content, which would be useful for identifying leadership in specific topics. Future research could address these areas.

	Our findings also reveal that interactions in online legislative networks are influenced by systemic network properties and institutional traits. The ERGM shows that personal traits are not a major factor in explaining system configuration. Similarly, these attributes do not define the patterns of SBM-induced communities. This may be due to the low representation of some individual attribute categories, such as the 1.9\% of Congress members identifying as LGBTIQ+. Although our results do not support homophily effects based on individual attributes, we stress that further investigation into the role of personal traits in online legislative network connections is needed, particularly regarding gender identity and sexual orientation, which is a recent and promising area in political science \citep[][]{hemphill2021drives, ayoub2022not}.

	Our results regarding the influence of party affiliation on network formation are consistent with contemporary literature, which suggests that Congress members build networks based on party membership \citep[][]{wojcik2019legislative}. However, we also identify that the chamber is a significant institutional factor in link formation. This finding contributes to research examining behavioral differences on Twitter between chambers \citep[][]{hemphill2021drives}. We argue that legislators tend to connect with members of the same chamber, even if they are from different parties. While the ERGM shows similar effects of party and chamber on connection probabilities, the SBM reveals communities of the same chamber with varying party affiliations. We encourage readers to explore which of these institutional attributes has a greater impact on legislative link formation in online social networks.

	Furthermore, including network systemic properties in the ERGM significantly improves model performance according to information criteria. Structural characteristics of the network have a statistically significant effect on system configuration. Our findings, within the context of online legislative networks, align with previous studies on co-sponsorship networks, which indicate that structural characteristics are crucial for modeling the system. Ignoring these characteristics can lead to misleading conclusions about the effects of nodal covariates on connection probabilities within the system \citep{cranmer2011inferential}.

	The importance of systemic network properties in modeling reveals that legislative links on social media platforms like Twitter follow not only homophily patterns but also equivalence patterns, where individuals connect based on their roles in the system \citep[][]{kolaczyk2014statistical}. Previous studies suggest that leaders are more likely to share connections with other leaders than with non-leaders \citep{wojcik2019legislative}. Our results confirm this observation and provide empirical evidence on the influence of structural characteristics on interactions between communities. Specifically, we find that communities with higher probabilities of intra-group interaction exhibit high average values in structural characteristics, while those with lower values are less likely to interact with other communities. Our analysis of structural characteristics by community highlights internal party divisions, reflecting hierarchies among their members. For example, communities 2 and 9 are dominant among Republicans and Democrats, respectively. Their high probability of interaction and prominent role in the system structure suggest that these communities are more likely to adopt extreme and politically polarized stances within each party \citep[][]{del2018echo}.

	Finally, it is important to note that our results are based on the 117th U.S. Congress and cross-sectional data. Future studies could extend this analysis to other Congresses and explore extensions of the ERGM and SBM or other models incorporating stochastic covariates and social connections in longitudinal data \citep[][]{hafiene2020influential} to capture changes and trends in the factors determining relationships and leadership in online legislative networks \citep{wang2024understanding}.

	\bibliographystyle{apalike}
	\bibliography{ref}

\begin{thebibliography}{}

\bibitem[Abbe, 2018]{abbe2018community}
Abbe, E. (2018).
\newblock Community detection and stochastic block models: recent developments.
\newblock {\em Journal of Machine Learning Research}, 18(177):1--86.

\bibitem[Ayoub, 2022]{ayoub2022not}
Ayoub, P.~M. (2022).
\newblock Not that niche: making room for the study of {LGBTIQ} people in
  political science.
\newblock {\em European Journal of Politics and Gender}, 5(2):154--172.

\bibitem[Bergersen et~al., 2018]{bergersen2018intersectionality}
Bergersen, M., Klar, S., and Schmitt, E. (2018).
\newblock Intersectionality and engagement among the lgbtq+ community.
\newblock {\em Journal of Women, Politics \& Policy}, 39(2):196--219.

\bibitem[Biernacki et~al., 1998]{biernacki1998assessing}
Biernacki, C., Celeux, G., and Govaert, G. (1998).
\newblock {\em Assessing a mixture model for clustering with the integrated
  classification likelihood}.
\newblock PhD thesis, INRIA.

\bibitem[Binder, 2022]{binder2022}
Binder, S. (2022).
\newblock Goodbye to the 117th {C}ongress, bookended by remarkable events.
  {T}he {W}ashington {P}ost. {A}vailable online:.
\newblock
  \url{https://www.washingtonpost.com/politics/2022/12/29/congress-year-review/}.
\newblock Accessed: 2024-07-29.

\bibitem[Bivens, 2023]{bivens2023federal}
Bivens, B.~M. (2023).
\newblock Federal care policy possibilities in the 117th congress: Toward
  expansive kinship and collectivized carework.
\newblock {\em Educational Studies}, 59(2):145--162.

\bibitem[Bollob{\'a}s et~al., 2001]{bollobas2001degree}
Bollob{\'a}s, B.~e., Riordan, O., Spencer, J., and Tusn{\'a}dy, G. (2001).
\newblock The degree sequence of a scale-free random graph process.
\newblock {\em Random Structures \& Algorithms}, 18(3):279--290.

\bibitem[Bond, 2024]{bond2024contemporary}
Bond, J.~R. (2024).
\newblock The contemporary presidency: Which presidents win more or less than
  expected in congress? a biden update.
\newblock {\em Presidential Studies Quarterly}, 54(2):271--283.

\bibitem[Borge~Bravo and Esteve Del~Valle, 2017]{borge2017opinion}
Borge~Bravo, R. and Esteve Del~Valle, M. (2017).
\newblock Opinion leadership in parliamentary twitter networks: A matter of
  layers of interaction?
\newblock {\em Journal of Information Technology \& Politics}, 14(3):263--276.

\bibitem[Bouchitt{\'e} and Todinca, 2002]{bouchitte2002listing}
Bouchitt{\'e}, V. and Todinca, I. (2002).
\newblock Listing all potential maximal cliques of a graph.
\newblock {\em Theoretical Computer Science}, 276(1-2):17--32.

\bibitem[Chakrabarti and Ghosh, 2011]{chakrabarti2011aic}
Chakrabarti, A. and Ghosh, J.~K. (2011).
\newblock Aic, bic and recent advances in model selection.
\newblock {\em Philosophy of statistics}, pages 583--605.

\bibitem[Cherepnalkoski and Mozeti{\v{c}}, 2016]{cherepnalkoski2016retweet}
Cherepnalkoski, D. and Mozeti{\v{c}}, I. (2016).
\newblock Retweet networks of the european parliament: Evaluation of the
  community structure.
\newblock {\em Applied network science}, 1:1--20.

\bibitem[Clemens and Veuger, 2021]{clemens2021politics}
Clemens, J. and Veuger, S. (2021).
\newblock Politics and the distribution of federal funds: Evidence from federal
  legislation in response to covid-19.
\newblock {\em Journal of Public Economics}, 204:104554.

\bibitem[Congress, 2021]{H3807}
Congress, U.~S. (2021).
\newblock Congressional record - {H}ouse {H}3807.
\newblock 07/24/2024.

\bibitem[Congress, 2022]{congress}
Congress, U.~S. (2022).
\newblock Proceedings and debates of the 117th {C}ongress.
\newblock 07/24/2024.

\bibitem[Cook, 2016]{cook2016american}
Cook, J.~M. (2016).
\newblock Are american politicians as partisan online as they are offline?
  twitter networks in the us senate and maine state legislature.
\newblock {\em Policy \& Internet}, 8(1):55--71.

\bibitem[Cranmer and Desmarais, 2011]{cranmer2011inferential}
Cranmer, S.~J. and Desmarais, B.~A. (2011).
\newblock Inferential network analysis with exponential random graph models.
\newblock {\em Political analysis}, 19(1):66--86.

\bibitem[Csardi, 2021]{csardi2013package}
Csardi, M.~G. (2021).
\newblock Package ‘igraph’.
\newblock {\em Last accessed}, 3(09):2013.

\bibitem[Danon et~al., 2005]{danon2005comparing}
Danon, L., Diaz-Guilera, A., Duch, J., and Arenas, A. (2005).
\newblock Comparing community structure identification.
\newblock {\em Journal of statistical mechanics: Theory and experiment},
  2005(09):P09008.

\bibitem[Davies, 2011]{davies2011legislative}
Davies, E. (2011).
\newblock Legislative update: Families beyond bars act of 2010.
\newblock {\em Child. Legal Rts. J.}, 31:61.

\bibitem[Del~Valle and Bravo, 2018]{del2018echo}
Del~Valle, M.~E. and Bravo, R.~B. (2018).
\newblock Echo chambers in parliamentary twitter networks: The catalan case.
\newblock {\em International journal of communication}, 12:1715--1735.

\bibitem[Eddington, 2018]{eddington2018communicative}
Eddington, S.~M. (2018).
\newblock The communicative constitution of hate organizations online: A
  semantic network analysis of “make america great again”.
\newblock {\em Social Media+ Society}, 4(3):2056305118790763.

\bibitem[Fink et~al., 2023]{fink2023congressional}
Fink, C.~G., Omodt, N., Zinnecker, S., and Sprint, G. (2023).
\newblock A congressional twitter network dataset quantifying pairwise
  probability of influence.
\newblock {\em Data in Brief}, 50:109521.

\bibitem[Frank and Strauss, 1986]{frank1986markov}
Frank, O. and Strauss, D. (1986).
\newblock Markov graphs.
\newblock {\em Journal of the american Statistical association},
  81(395):832--842.

\bibitem[Friedkin, 1981]{friedkin1981development}
Friedkin, N.~E. (1981).
\newblock The development of structure in random networks: an analysis of the
  effects of increasing network density on five measures of structure.
\newblock {\em Social Networks}, 3(1):41--52.

\bibitem[Gamm and Huber, 2002]{gamm2002legislatures}
Gamm, G. and Huber, J. (2002).
\newblock Legislatures as political institutions: Beyond the contemporary
  congress.
\newblock {\em Political science: State of the discipline}, pages 313--43.

\bibitem[Geyer and Thompson, 1992]{geyer1992constrained}
Geyer, C.~J. and Thompson, E.~A. (1992).
\newblock Constrained monte carlo maximum likelihood for dependent data.
\newblock {\em Journal of the Royal Statistical Society: Series B
  (Methodological)}, 54(3):657--683.

\bibitem[Gjoka et~al., 2014]{gjoka2014estimating}
Gjoka, M., Smith, E., and Butts, C. (2014).
\newblock Estimating clique composition and size distributions from sampled
  network data.
\newblock In {\em 2014 IEEE Conference on Computer Communications Workshops
  (INFOCOM WKSHPS)}, pages 837--842. IEEE.

\bibitem[Golbeck et~al., 2010]{golbeck2010twitter}
Golbeck, J., Grimes, J.~M., and Rogers, A. (2010).
\newblock Twitter use by the us congress.
\newblock {\em Journal of the American society for information science and
  technology}, 61(8):1612--1621.

\bibitem[Grant et~al., 2010]{grant2010digital}
Grant, W.~J., Moon, B., and Busby~Grant, J. (2010).
\newblock Digital dialogue? australian politicians' use of the social network
  tool twitter.
\newblock {\em Australian journal of political science}, 45(4):579--604.

\bibitem[Green, 2019]{green2019legislative}
Green, M. (2019).
\newblock {\em Legislative hardball: The House Freedom Caucus and the power of
  threat-making in Congress}.
\newblock Cambridge University Press.

\bibitem[Hafiene et~al., 2020]{hafiene2020influential}
Hafiene, N., Karoui, W., and Romdhane, L.~B. (2020).
\newblock Influential nodes detection in dynamic social networks: A survey.
\newblock {\em Expert Systems with Applications}, 159:113642.

\bibitem[Hemphill et~al., 2021]{hemphill2021drives}
Hemphill, L., Russell, A., and Sch{\"o}pke-Gonzalez, A.~M. (2021).
\newblock What drives us congressional members’ policy attention on twitter?
\newblock {\em Policy \& Internet}, 13(2):233--256.

\bibitem[Himelboim et~al., 2017]{himelboim2017classifying}
Himelboim, I., Smith, M.~A., Rainie, L., Shneiderman, B., and Espina, C.
  (2017).
\newblock Classifying twitter topic-networks using social network analysis.
\newblock {\em Social media+ society}, 3(1):2056305117691545.

\bibitem[Hubert and Arabie, 1985]{hubert1985comparing}
Hubert, L. and Arabie, P. (1985).
\newblock Comparing partitions.
\newblock {\em Journal of classification}, 2:193--218.

\bibitem[Hunter and Handcock, 2006]{hunter2006inference}
Hunter, D.~R. and Handcock, M.~S. (2006).
\newblock Inference in curved exponential family models for networks.
\newblock {\em Journal of computational and graphical statistics},
  15(3):565--583.

\bibitem[Hunter et~al., 2008]{hunter2008ergm}
Hunter, D.~R., Handcock, M.~S., Butts, C.~T., Goodreau, S.~M., and Morris, M.
  (2008).
\newblock ergm: A package to fit, simulate and diagnose exponential-family
  models for networks.
\newblock {\em Journal of statistical software}, 24(3):nihpa54860.

\bibitem[Kirkland and Gross, 2014]{kirkland2014measurement}
Kirkland, J.~H. and Gross, J.~H. (2014).
\newblock Measurement and theory in legislative networks: The evolving topology
  of congressional collaboration.
\newblock {\em Social networks}, 36:97--109.

\bibitem[Kleinberg, 1999]{kleinberg1999authoritative}
Kleinberg, J.~M. (1999).
\newblock Authoritative sources in a hyperlinked environment.
\newblock {\em Journal of the ACM (JACM)}, 46(5):604--632.

\bibitem[Kolaczyk and Cs{\'a}rdi, 2014]{kolaczyk2014statistical}
Kolaczyk, E.~D. and Cs{\'a}rdi, G. (2014).
\newblock {\em Statistical analysis of network data with R}, volume~65.
\newblock Springer.

\bibitem[Lawson, 2021]{lawson2021clean}
Lawson, A.~J. (2021).
\newblock Clean energy standards: Selected issues for the 117th congress.
\newblock {\em Congressional Research Service (CRS) Reports and Issue Briefs},
  pages NA--NA.

\bibitem[Lee and Wilkinson, 2019]{lee2019review}
Lee, C. and Wilkinson, D.~J. (2019).
\newblock A review of stochastic block models and extensions for graph
  clustering.
\newblock {\em Applied Network Science}, 4(1):1--50.

\bibitem[Leger, 2016]{leger2016blockmodels}
Leger, J.-B. (2016).
\newblock Blockmodels: A r-package for estimating in latent block model and
  stochastic block model, with various probability functions, with or without
  covariates.
\newblock {\em arXiv preprint arXiv:1602.07587}.

\bibitem[Leger et~al., 2021]{leger2022}
Leger, J.-B., Barbillon, P., and Chiquet, J. (2021).
\newblock {\em blockmodels: Latent and Stochastic Block Model Estimation by a
  'V-EM' Algorithm}.
\newblock R package version 1.1.5.

\bibitem[Li et~al., 2021]{li2021review}
Li, N., Huang, Q., Ge, X., He, M., Cui, S., Huang, P., Li, S., and Fung, S.-F.
  (2021).
\newblock A review of the research progress of social network structure.
\newblock {\em Complexity}, 2021(1):6692210.

\bibitem[Lomet et~al., 2012]{lomet2012model}
Lomet, A., Govaert, G., and Grandvalet, Y. (2012).
\newblock Model selection in block clustering by the integrated classification
  likelihood.
\newblock In {\em 20th International Conference on Computational Statistics
  (COMPSTAT 2012)}, pages 519--530.

\bibitem[Luque and Sosa, 2022]{luque2022operationalizing}
Luque, C. and Sosa, J. (2022).
\newblock Operationalizing legislative bodies: A methodological and empirical
  perspective.
\newblock {\em arXiv preprint arXiv:2211.17066}.

\bibitem[Lusher et~al., 2013]{lusher2013exponential}
Lusher, D., Koskinen, J., and Robins, G. (2013).
\newblock {\em Exponential random graph models for social networks: Theory,
  methods, and applications}.
\newblock Cambridge University Press.

\bibitem[Mamet, 2021]{mamet2021representation}
Mamet, E. (2021).
\newblock Representation on the periphery: The past and future of nonvoting
  members of congress.
\newblock {\em American Political Thought}, 10(3):390--418.

\bibitem[Mankad and Michailidis, 2015]{mankad2015analysis}
Mankad, S. and Michailidis, G. (2015).
\newblock Analysis of multiview legislative networks with structured matrix
  factorization: Does twitter influence translate to the real world?
\newblock {\em The Annals of Applied Statistics}, pages 1950--1972.

\bibitem[Marcaurelle, 1999]{marcaurelle1999payday}
Marcaurelle, M. (1999).
\newblock Payday lending practices spark federal legislation.
\newblock {\em Pub. Int. L. Rep.}, 4:7.

\bibitem[Morris et~al., 2008]{morris2008specification}
Morris, M., Handcock, M.~S., and Hunter, D.~R. (2008).
\newblock Specification of exponential-family random graph models: terms and
  computational aspects.
\newblock {\em Journal of statistical software}, 24(4):1548.

\bibitem[Newman, 2002]{newman2002assortative}
Newman, M.~E. (2002).
\newblock Assortative mixing in networks.
\newblock {\em Physical review letters}, 89(20):208701.

\bibitem[Noldus and Van~Mieghem, 2015]{noldus2015assortativity}
Noldus, R. and Van~Mieghem, P. (2015).
\newblock Assortativity in complex networks.
\newblock {\em Journal of Complex Networks}, 3(4):507--542.

\bibitem[Pearson, 2022]{pearson2022legacies}
Pearson, A. (2022).
\newblock The legacies of trump’s battles with congress.
\newblock {\em The Trump effect: Disruption and its consequences in US politics
  and government}, pages 65--82.

\bibitem[Praet et~al., 2021]{praet2021patterns}
Praet, S., Martens, D., and Van~Aelst, P. (2021).
\newblock Patterns of democracy? social network analysis of parliamentary
  twitter networks in 12 countries.
\newblock {\em Online Social Networks and Media}, 24:100154.

\bibitem[Rakhmawati and Mufidah, 2020]{rakhmawati2020social}
Rakhmawati, N.~A. and Mufidah, K. (2020).
\newblock Social network analysis of legislative candidates in indonesia
  general election 2019 using community detection.
\newblock In {\em 2020 3rd International Conference on Computer and Informatics
  Engineering (IC2IE)}, pages 306--310. IEEE.

\bibitem[Rand, 1971]{rand1971objective}
Rand, W.~M. (1971).
\newblock Objective criteria for the evaluation of clustering methods.
\newblock {\em Journal of the American Statistical association},
  66(336):846--850.

\bibitem[Ringe and Wilson, 2016]{ringe2016pinpointing}
Ringe, N. and Wilson, S.~L. (2016).
\newblock Pinpointing the powerful: Covoting network centrality as a measure of
  political influence.
\newblock {\em Legislative Studies Quarterly}, 41(3):739--769.

\bibitem[Rossiter et~al., 2018]{rossiter2018congressional}
Rossiter, K.~M., Wong, D.~W., and Delamater, P.~L. (2018).
\newblock Congressional redistricting: Keeping communities together?
\newblock {\em The Professional Geographer}, 70(4):609--623.

\bibitem[Schwarzenbach and Jensen, 2024]{schwarzenbach2024extremists}
Schwarzenbach, A. and Jensen, M. (2024).
\newblock Extremists of a feather flock together? community structures,
  transitivity, and patterns of homophily in the us islamist co-offending
  network.
\newblock {\em PLoS one}, 19(6):e0298273.

\bibitem[Van~Vliet et~al., 2020]{van2020twitter}
Van~Vliet, L., T{\"o}rnberg, P., and Uitermark, J. (2020).
\newblock The twitter parliamentarian database: Analyzing twitter politics
  across 26 countries.
\newblock {\em PLoS one}, 15(9):e0237073.

\bibitem[Vergeer, 2015]{vergeer2015twitter}
Vergeer, M. (2015).
\newblock Twitter and political campaigning.
\newblock {\em Sociology compass}, 9(9):745--760.

\bibitem[Wang et~al., 2024]{wang2024understanding}
Wang, Z., Fellows, I.~E., and Handcock, M.~S. (2024).
\newblock Understanding networks with exponential-family random network models.
\newblock {\em Social Networks}, 78:81--91.

\bibitem[Ward et~al., 2011]{ward2011network}
Ward, M.~D., Stovel, K., and Sacks, A. (2011).
\newblock Network analysis and political science.
\newblock {\em Annual Review of Political Science}, 14:245--264.

\bibitem[Wasserman and Pattison, 1996]{wasserman1996logit}
Wasserman, S. and Pattison, P. (1996).
\newblock Logit models and logistic regressions for social networks: I. an
  introduction to markov graphs and p*.
\newblock {\em Psychometrika}, 61(3):401--425.

\bibitem[Wojcik, 2019]{wojcik2019legislative}
Wojcik, S. (2019).
\newblock Why legislative networks? analyzing legislative network formation.
\newblock {\em Political Science Research and Methods}, 7(3):505--522.

\bibitem[Zhang and Batjargal, 2022]{zhang2022review}
Zhang, M. and Batjargal, T. (2022).
\newblock Review on new spending of united states bipartisan infrastructure
  bill.
\newblock {\em Journal of Infrastructure, Policy and Development}, 6(2):1507.

\end{thebibliography}

\end{document}